\documentclass[aps,prb,twocolumn,floatfix]{revtex4-1}
\usepackage{times} 
\usepackage{epsfig} 
\usepackage{amsmath}
\usepackage{amssymb} 
\usepackage{graphicx}
\usepackage{dsfont}
\usepackage{bbm}
\usepackage[colorlinks=true,citecolor=red,linkcolor=blue]{hyperref}

\newcommand{\tsh}{t_\mathrm{h}}

\newcommand{\bfk}{\mathbf{k}}

\newcommand{\bfA}{\mathbf{A}}

\begin{document}

\title{Floquet Hofstadter Butterfly on the Kagome and Triangular Lattices}
\author{Liang Du, Qi Chen, Aaron D. Barr, Ariel R. Barr, Gregory A. Fiete }
\affiliation{Department of Physics, The University of Texas at Austin, Austin, TX 78712, USA}

\begin{abstract}
In this work we use Floquet theory to theoretically study the influence of monochromatic circularly and linearly polarized light on the Hofstadter butterfly---induced by a uniform perpendicular magnetic field--for both the kagome and triangular lattices. In the absence of the laser light, the butterfly has fractal structure with inversion symmetry about magnetic flux $\phi=1/4$, and reflection symmetry about $\phi=1/2$. 
As the system is exposed to an external laser, we find circularly polarized light deforms the butterfly by breaking the mirror symmetry at flux $\phi=1/2$.  By contrast, linearly polarized light deforms the original butterfly while preserving the mirror symmetry at flux $\phi=1/2$.  We find the inversion symmetry is always preserved for both linear and circular polarized light. For linearly polarized light, the Hofstadter butterfly depends on the polarization direction.  Further, we study the effect of the laser on the Chern number of lowest band in the off-resonance regime (laser frequency is larger than the bandwidth). For circularly polarized light, we find that low laser intensity will not change the Chern number, but beyond a critical intensity the Chern number will change. For linearly polarized light, the Chern number depends on the polarization direction.  Our work highlights the generic features expected for the periodically driven Hofstadter problem on different lattices.
\end{abstract}
\date{\today}
\maketitle

\section{INTRODUCTION}
The Hofstadter butterfly--the energy spectrum of a two-dimensional lattice model as a function of static magnetic flux through the unit cell--exhibits a complex fractal structure resembling a butterfly.\cite{Hofstadter:prb76} The original butterfly was based on a tight-binding model for the two-dimensional square lattice. Subsequent work generalized the square lattice result to triangular, honeycomb, and kagome lattices.\cite{LiJ:jpcm11} Even through there exist some differences in detail, the fractal pattern is observed for all of the above lattices.  On the square or honeycomb lattice with an isotropic hopping parameter,  the system exhibits particle-hole symmetry, which makes the Hofstadter butterfly symmetric about the zero-energy axis. Further, a reflection symmetry about $1/2$ flux (in units of the fundamental flux quantum $hc/e$ where $h$ is Planck's constant, $c$ is the speed of light, and $e$ is the charger of the electron) is observed. In the triangular and kagome lattices, the particle-hole symmetry is broken, and the reflection symmetry about the zero-energy axis disappears, while the reflection about the $1/2$ flux axis is preserved. Moreover, an additional central (inversion) symmetry about the point with zero energy and $1/4$ flux is observed. 

The strength of the magnetic field required to observe the Hofstadter butterfly depends on the spacing between atoms in the lattice (i.e., the lattice constant).\cite{Hofstadter:prb76}  For conventional materials, the magnitude of the magnetic field required to observe the fractal pattern is on the order of $10^4$ Tesla, well above the field generated by the best magnets currently available (about 100 Tesla). 

One way to circumvent this problem is to use artificial superlattices, where the lattice spacing can be an order of magnitude larger than in conventional materials.  In 1998, the Hofstadter butterfly was reproduced in experiments with microwaves transmitted through a waveguide equipped with an array of scatterers.\cite{Kuhl:prl98}  In 2013, several experimental groups independently reported evidence of the Hofstadter butterfly spectrum in graphene devices fabricated on hexagonal boron nitride substrates.\cite{Dean:nat13,Ponomarenko:nat13,Hunt:sci13} In 2017, a simulation of two-dimensional electrons in a magnetic field using interacting photons in nine superconducting qubits exhibited a Hofstadter butterfly.\cite{Roushan:sci17}

Recently, light-driven materials have attracted considerable interest from the physics community.  At the non-interacting level, dramatic changes in the band structure can occur, including a change from a non-topological band structure to a topological one.\cite{Kitagawa:prb10,Rudner:prx13,Katan:prl13,Lindner:prb13,Dora:prl12,Inoue:prl10,Cayssol:pssr13,Kitagawa:prb11,Iadecola:prl13,Ezawa:prl13,Kemper:prb13,Rechtsman:nat13,DuL:prb17a,ChenQ:prb18}  
Two commonly discussed physical scenarios for periodically driven systems include periodic changes in the laser fields that establish the optical lattice potential for cold atom systems,\cite{Jotzu:nat14,Bilitewski:pra15} and solid state systems that are driven by a monochromatic laser field.\cite{Fregoso:prb13,Sentef:nc15,WangY:sci13,Mahmood:np16,Calvo:prb15,Lago:pra15,Perez-Piskunow:pra15, Perez-Piskunow:prb14} 

The effect of light on the Hofstadter butterfly has not been studied extensively. \cite{Lababidi:prl14, ZhouZ:prb14, Wackerl:arXiv18, Kooi:arXiv18} Prior work based on the square lattice Hofstadter model found that periodic driving leads to pairs of counter-propagating chiral edge modes, which are protected by the chiral symmetry and robust against static disorder.\cite{Lababidi:prl14, ZhouZ:prb14} In Ref.[\onlinecite{Wackerl:arXiv18}], the driven Hofstadter butterfly on the honeycomb lattice was studied under the influence of circularly and linearly polarized light, and the Chern number of the ``ground state" of the Floquet-Hofstadter spectrum was studied for an off-resonant laser.  Recently, the effect of a circularly polarized laser on the Hofstadter butterfly was studied systematically.\cite{Kooi:arXiv18}  By decreasing the laser frequency from the off-resonant to the on-resonant regime, the authors\cite{Kooi:arXiv18} found that the ``top" two bands do not hybridize with the bands of the ``upper" Floquet copy. The observed phenomena is well explained using an effective Hamiltonian. 

As stated before, the equilibrium Hofstadter butterfly is considerably different for the triangular or kagome lattices when compared to that of the square or honeycomb lattices. Previous out-of-equilibrium (periodically-driven) studies of the Hofstadter butterfly have focused on the square lattice or honeycomb lattice. In this work, we focus our attention on the effect of circularly and linearly polarized light on the Hofstadter butterfly--and the corresponding Chern numbers--for the kagome and triangular lattices. 

The organization of this paper is as follows. We study tight-binding Hamiltonians on a the kagome and triangular lattices exposed to a perpendicular magnetic field and monochromatic laser. In Sec.\ref{sec:model}, we introduce the model Hamiltonian on these two lattices. The effect of laser on the Hofstadter butterfly is studied systematically in Sec.\ref{sec:butterfly}, and the Chern number is calculated in Sec.\ref{sec:chern}. The numerical results for the triangular lattice are presented in Sec.\ref{sec:triangular}. Finally, in Sec.\ref{sec:conclusion}, we summarize our main results and conclusions.

\label{sec:intro}

\section{Model and method}
\label{sec:model}
\begin{figure}[t]
\includegraphics[width=0.99\linewidth, angle=0]{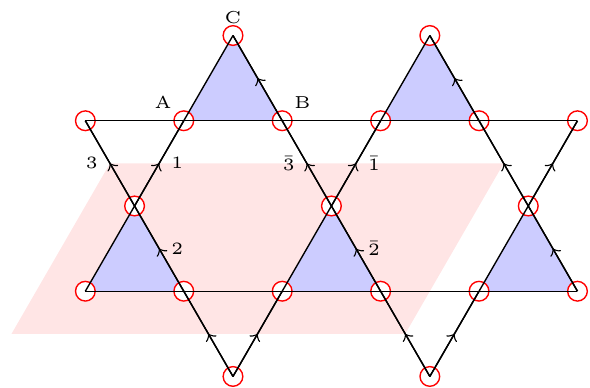}  %
\caption{(Color online.) The kagome lattice with the three (for the case of zero magnetic field) sites (A, B, C) in one unit cell are labeled. 
Three nearest-neighbor unit vectors are $\vec{\delta}_{1}=(1,0)a$, $\vec{\delta}_{2}=(1/2,\sqrt{3}/2)a$, and
$\vec{\delta}_{3}=\vec{\delta}_{2}-\vec{\delta}_{1} = (-1/2,\sqrt{3}/2)a$ with $a$ the nearest-neighbor distance in the kagome lattice. The translational vectors are $\vec{a}_1 = 2\vec{\delta}_1$ and $\vec{a}_2 = 2\vec{\delta}_2$. The reciprocal lattice vectors are $\vec{b}_1 = (1, -1/\sqrt{3})\pi/a$ and $\vec{b}_2 = (0, 2/\sqrt{3})\pi/a$. 
When the system is exposed to a perpendicular magnetic field, the magnetic unit cell must be enlarged (to recover the translational symmetry) by an amount that depends on the value of magnetic flux $\phi$. For example, the magnetic cell is the pink area (shaded parallelogram) for magnetic flux $\phi/\phi_0 = 1/8q$ with $q=2$ ($\phi_0$ is defined as magnetic flux quantum).}
\label{fig:magnetic-unitcell}
\end{figure}

The model Hamiltonian we study, defined on a two-dimensional triangular or kagome lattice, is based on the isotropic nearest-neighbor hopping model, 
\begin{equation}\label{eq:htbr}
     H = -t_{\text h} \sum_{\langle ij \rangle, \sigma} c_{i\sigma}^{\dagger} c_{j\sigma},
\end{equation}
where $t_{\text h}$ is the isotropic hopping integral between nearest neighbors, 
$c_{i\sigma}^\dagger$ ($c_{j\sigma}$) creates (annihilates) an electron with spin $\sigma$ on site $i$ ($j$) of the two-dimensional lattice, 
and $\langle ij \rangle$ limits the summation to nearest neighbors.

\subsection{Equilibrium Hamiltonian without magnetic field}
The three-band kagome lattice model we study is based on the nearest-neighbor hopping model, Eq.\eqref{eq:htbr}. The kagome lattice is two-dimensional corner-sharing network of triangles shown in Fig.\ref{fig:magnetic-unitcell}.
To make the translational symmetry apparent, the Hamiltonian in real-space can be rewritten (omitting the spin index for clarity),
\begin{align}
 H_{\rm kagome} =&\sum_{m,n} c^\dagger_{m,n} a^{}_{m,n} + c^\dagger_{m,n} a^{}_{m,n+1} + h.c. \nonumber\\
 +&\sum_{m,n} c^\dagger_{m,n} b^{}_{m,n} + c^\dagger_{m,n} b^{}_{m-1,n+1} + h.c.\nonumber\\
 +&\sum_{m,n} b^\dagger_{m,n} a^{}_{m,n} + b^\dagger_{m,n} a^{}_{m+1,n} + h.c.,
\end{align}
where we define the position of an arbitrary unit cell as
\begin{align}
R(m,n) = m \vec{a}_{1} + n\vec{a}_{2},
\label{eq:vector-original}
\end{align}
with $m,n$ are integers, $a_{m,n}, b_{m,n}, c_{m,n}$ define annihilate operators on the three basis sites $A, B, C$ in the triangular unit cell $R(m,n)$ shown in Fig.\ref{fig:magnetic-unitcell}. 

Fourier transforming to momentum space, the Hamiltonian becomes,
$H = \sum_{{\bf k}} \psi_{{\bfk}}^\dagger {H}_{\bf k} \psi_{{\bfk}}$ with 
$\psi_{{\bfk}}=(a_{{\bfk}}, b_{{\bfk}}, c_{{\bfk}})^T$,
\begin{equation}\label{eq:htbk}
        {H}_{\mathbf{k}} = - \tsh
    \begin{pmatrix}
        0 & 1+e^{-i k_1} & 1+e^{-i k_2}\\
        1+e^{+i k_1} & 0 & 1+e^{-i k_3}\\
        1+e^{+i k_2} & 1+e^{+i k_3} & 0
    \end{pmatrix},
\end{equation}
where we used $k_i=\bfk\cdot\vec{a}_i$. 
Setting the distance between nearest neighbors to be $1$, 
the nearest-neighbor vectors are $\vec{\delta}_{1}=(1,0)a$, $\vec{\delta}_{2}=(1/2,\sqrt{3}/2)a$, and
$\vec{\delta}_{3}=\vec{\delta}_{2}-\vec{\delta}_{1} = (-1/2,\sqrt{3}/2)a$ with $a$ the nearest-neighbor distance.  The translational lattice vectors are $\vec{a}_1 = 2\vec{\delta}_1 = (2, 0)a$ and $\vec{a}_2 = 2\vec{\delta}_2=(1,\sqrt{3})a$. 
\begin{figure*}[t]
\includegraphics[width=0.49\linewidth, angle=0]{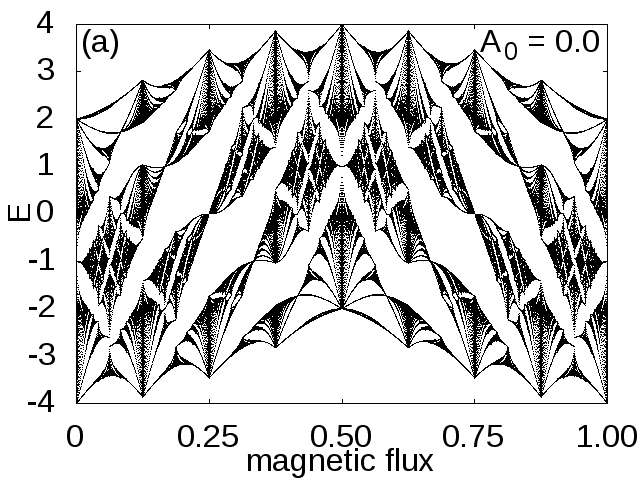}  %
\includegraphics[width=0.49\linewidth, angle=0]{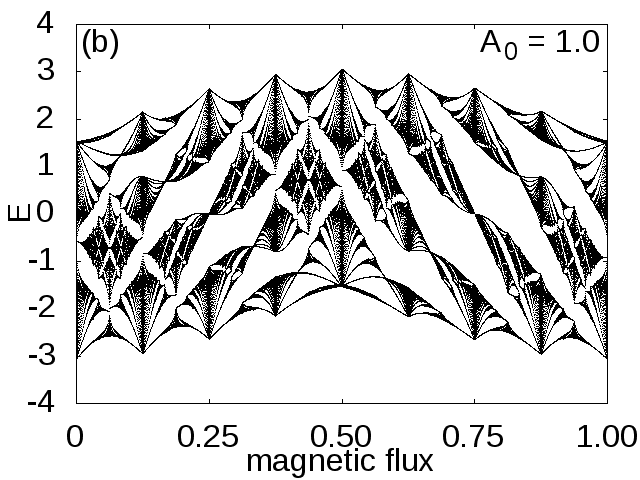}  %
\includegraphics[width=0.49\linewidth, angle=0]{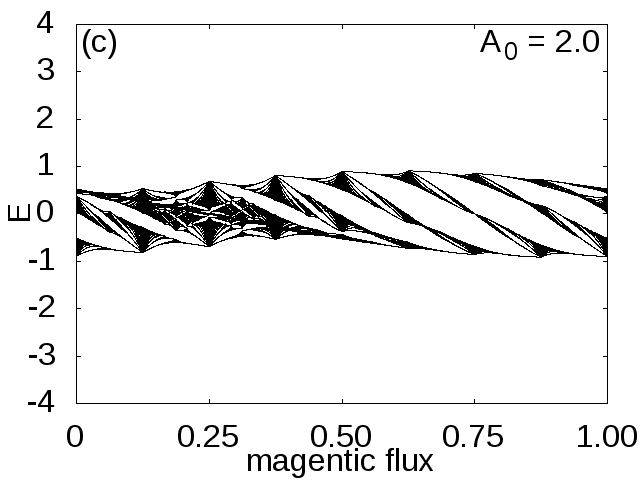}  %
\includegraphics[width=0.49\linewidth, angle=0]{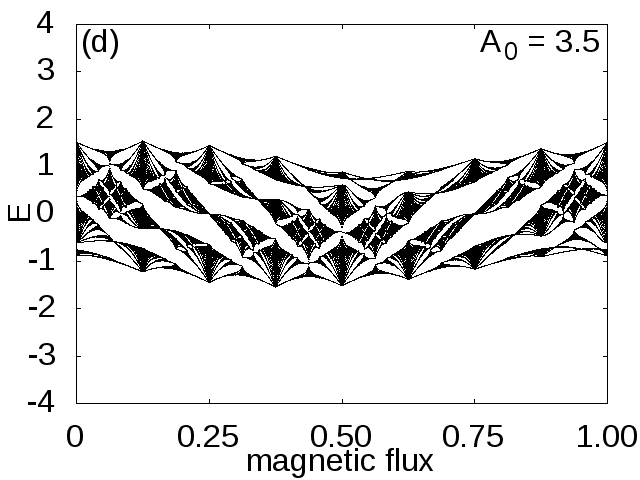}  %
\caption{(Color online.) The Hofstader butterfly for the kagome lattice, deformed by circularly polarized light with frequency fixed at an off-resonant regime $\Omega = 9.0$. The representative laser laser intensity are chosen as (a) $A_0 = 0.0$, (b) $A_0 = 1.0$, (c) $A_0 = 2.0$ and (d) $A_0 = 3.5$. The calculations are done with 5 Floquet copies. The magnetic flux is defined as $\phi = p/8q$ with p ranging from 1 to $8q-1$ and $q=199$.
}
\label{fig:kagome-fig1}
\end{figure*}
\begin{figure*}[t]
\includegraphics[width=0.49\linewidth, angle=0]{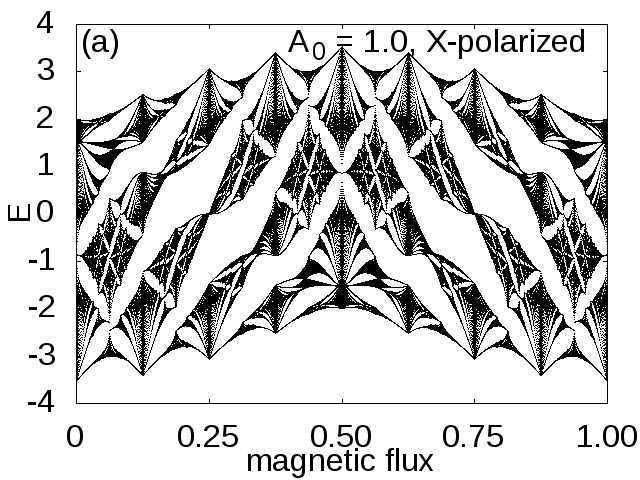}  %
\includegraphics[width=0.49\linewidth, angle=0]{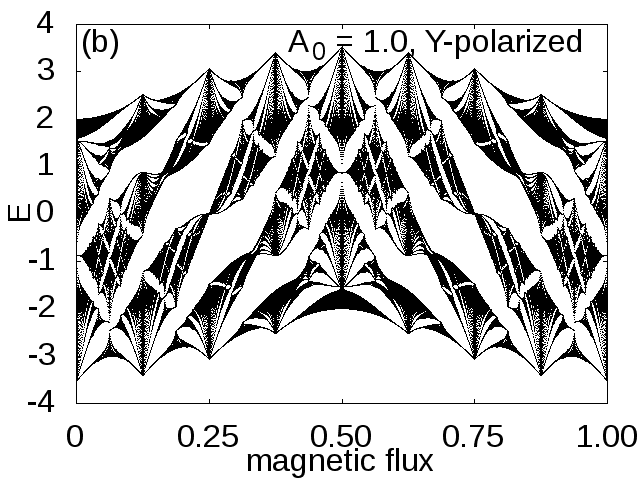}  %
\includegraphics[width=0.49\linewidth, angle=0]{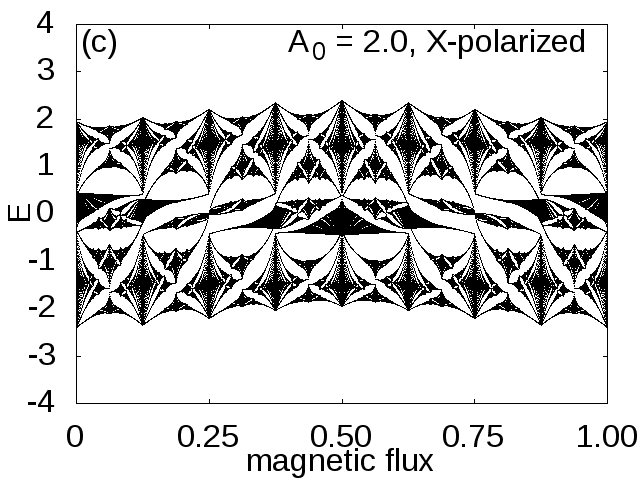}  %
\includegraphics[width=0.49\linewidth, angle=0]{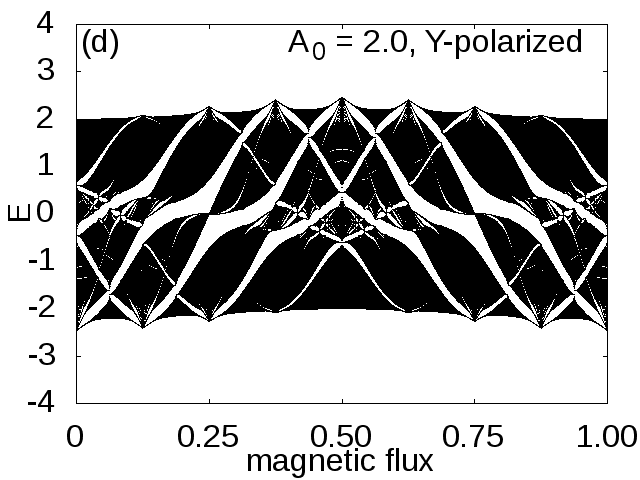}  %
\caption{(Color online) Hofstader's butterfly for the Kagome lattice exposed to linearly polarized laser with vector potential $A(t) = A_0 \sin(\Omega t)(\cos\alpha, \sin\alpha)$ where laser frequency is fixed at off-resonance region $\omega = 9.0$. (a) $A_0 = 1.0$, $\alpha = 0$;
(b) $A_0 = 1.0$, $\alpha = \pi/2$; (c) $A_0 = 2.0$, $\alpha = 0$; (d) $A_0 = 2.0$, $\alpha = \pi/2$. The calculations are done with 5 Floquet copies. The magnetic flux is defined as $\phi = p/8q$ with p ranging from 1 to $8q-1$ and $q=199$.
}
\label{fig:kagome-fig2}
\end{figure*}
\begin{figure*}[t]
\includegraphics[width=0.49\linewidth, angle=0]{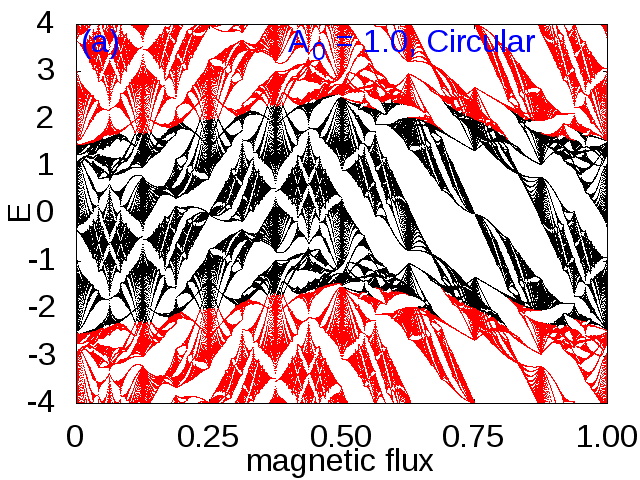}  %
\includegraphics[width=0.49\linewidth, angle=0]{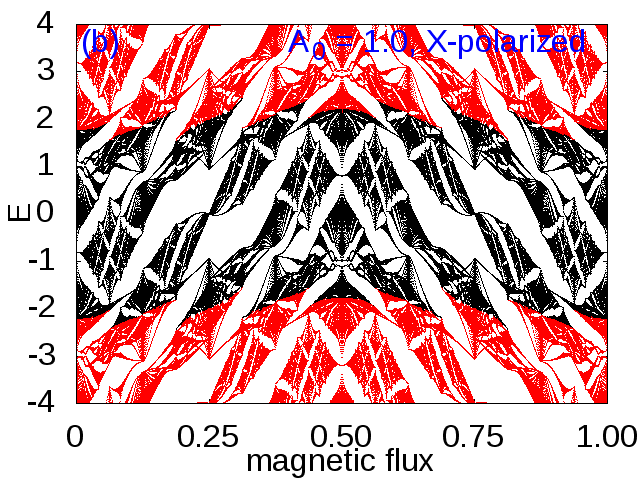}  %
\includegraphics[width=0.49\linewidth, angle=0]{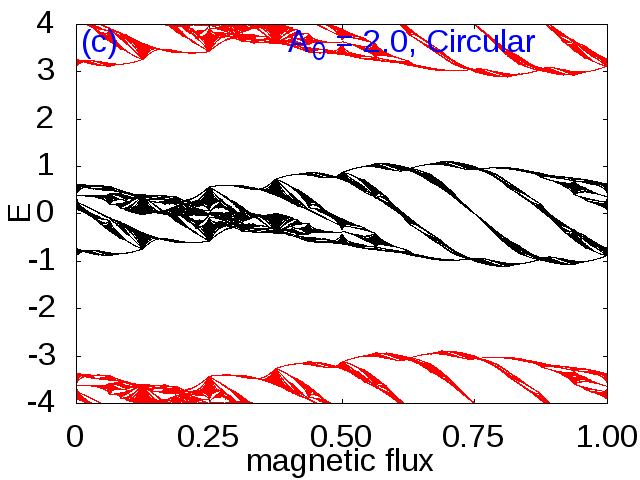}  %
\includegraphics[width=0.49\linewidth, angle=0]{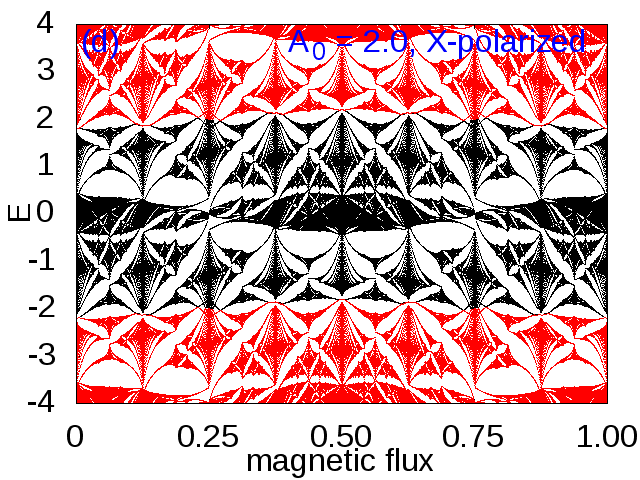}  %
\caption{(Color online.) The Hofstader butterfly for the kagome lattice exposed to circularly (a,c) and linearly (b,d) polarized lasers with the laser frequency fixed at on-resonance value $\Omega = 4.0$. (a) circularly polarized light $A_0 = 1.0$.
(b) linearly polarized light $A_0 = 1.0$, $\alpha = 0$. (c) circularly polarized light $A_0 = 2.0$. (d) linearly polarized light $A_0 = 2.0$, $\alpha = 0$. The red points are energy spectrum from upper and lower Floquet copies. The calculations are done with 9 Floquet copies. The magnetic flux is defined as $\phi = p/8q$ with p ranging from 1 to $8q-1$ and $q=199$.
}
\label{fig:kagome-fig3}
\end{figure*}

\begin{figure*}[t]
\includegraphics[width=0.49\linewidth, angle=0]{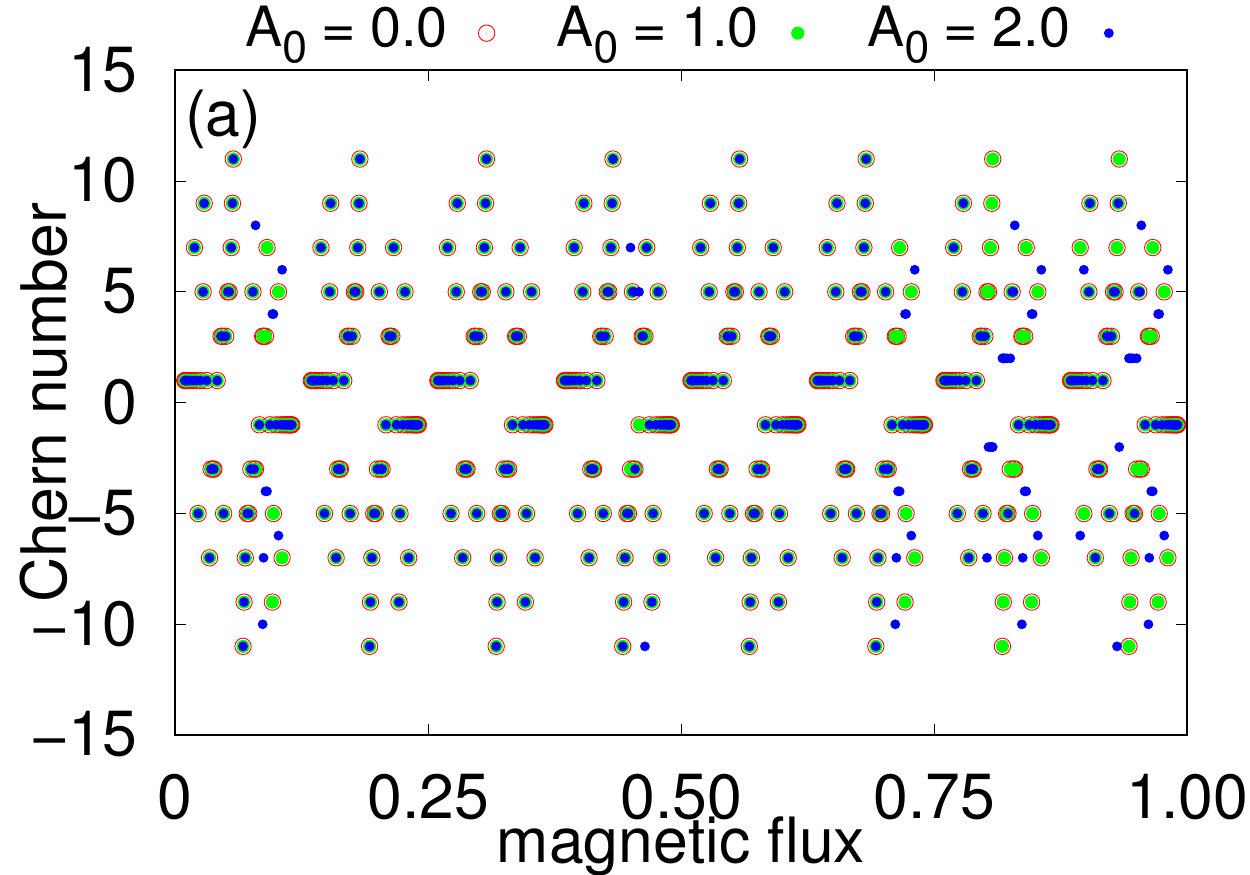}  %
\includegraphics[width=0.49\linewidth, angle=0]{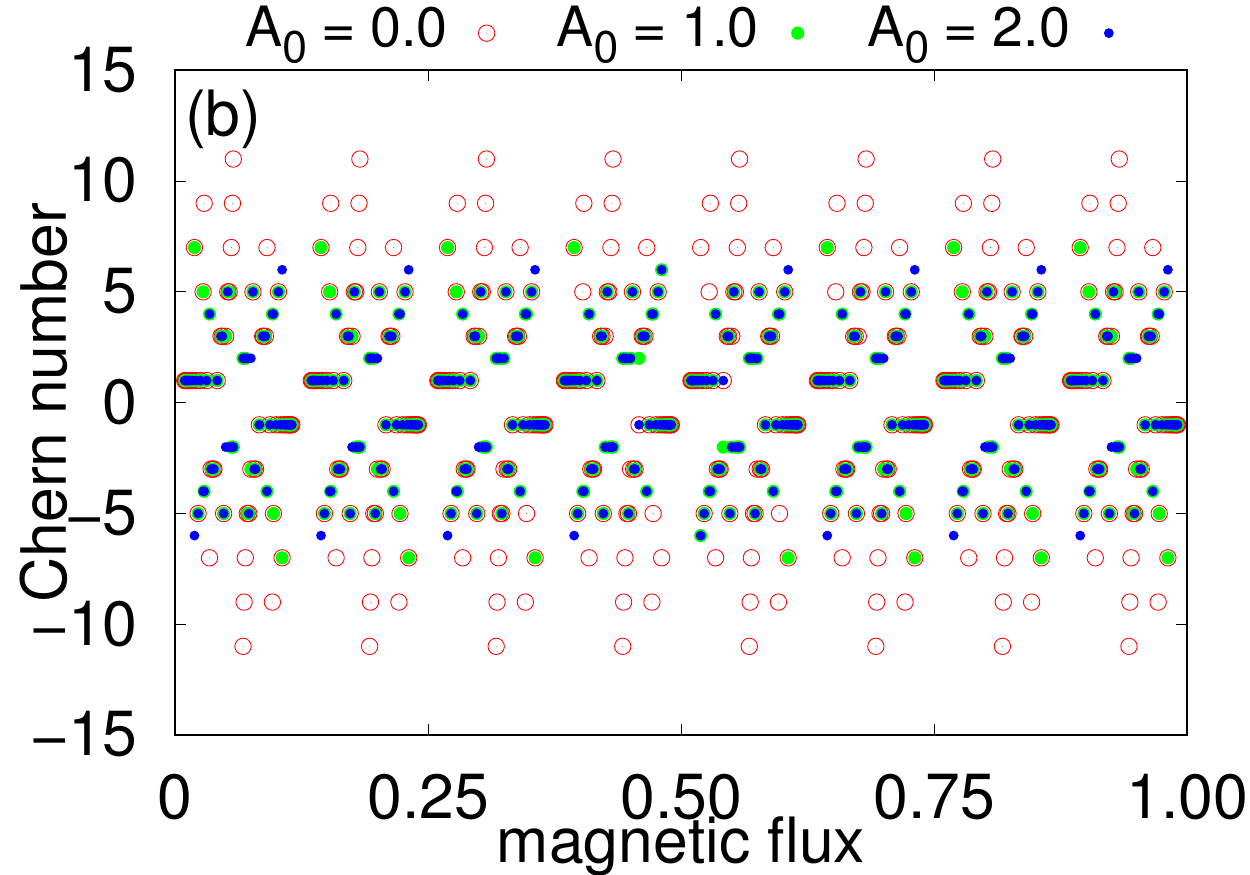}  %
\includegraphics[width=0.49\linewidth, angle=0]{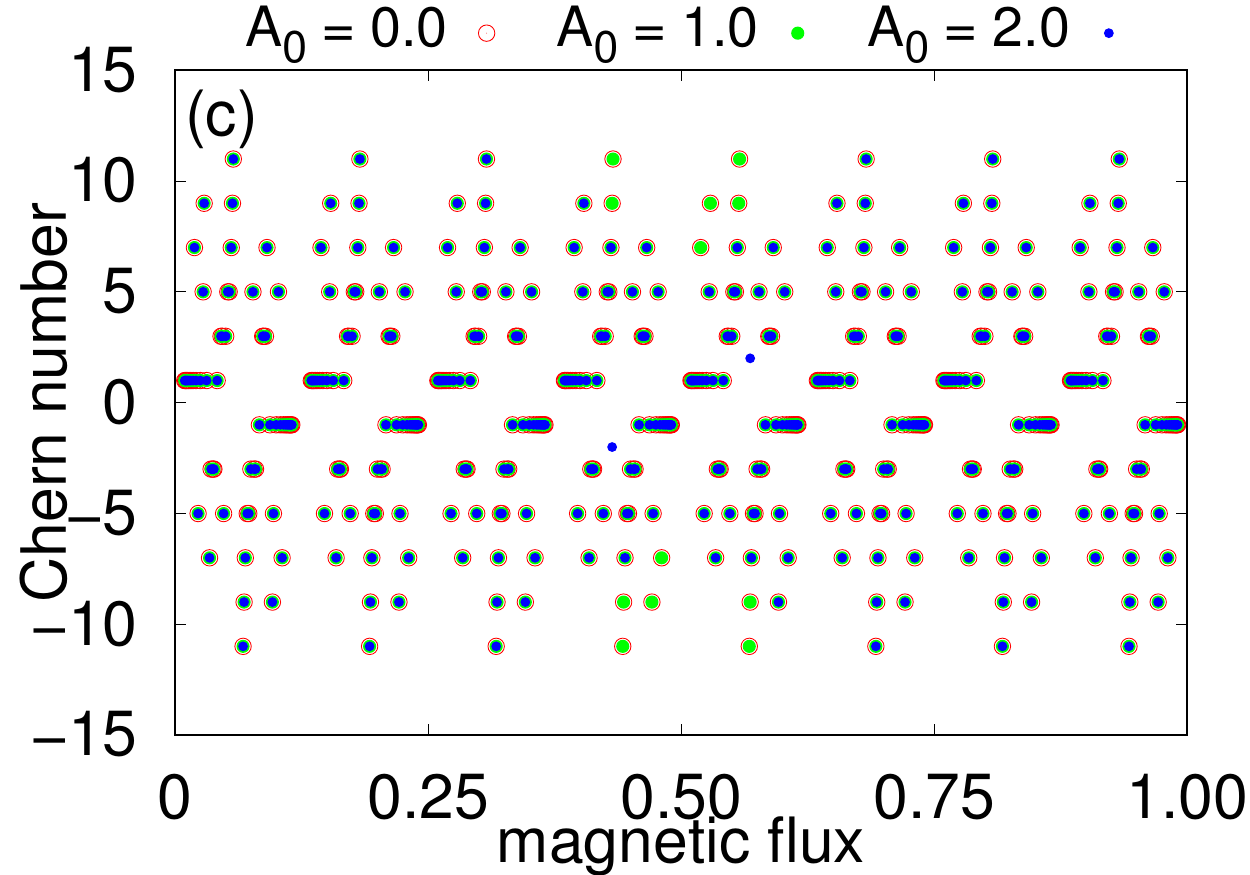}  %
\includegraphics[width=0.49\linewidth, angle=0]{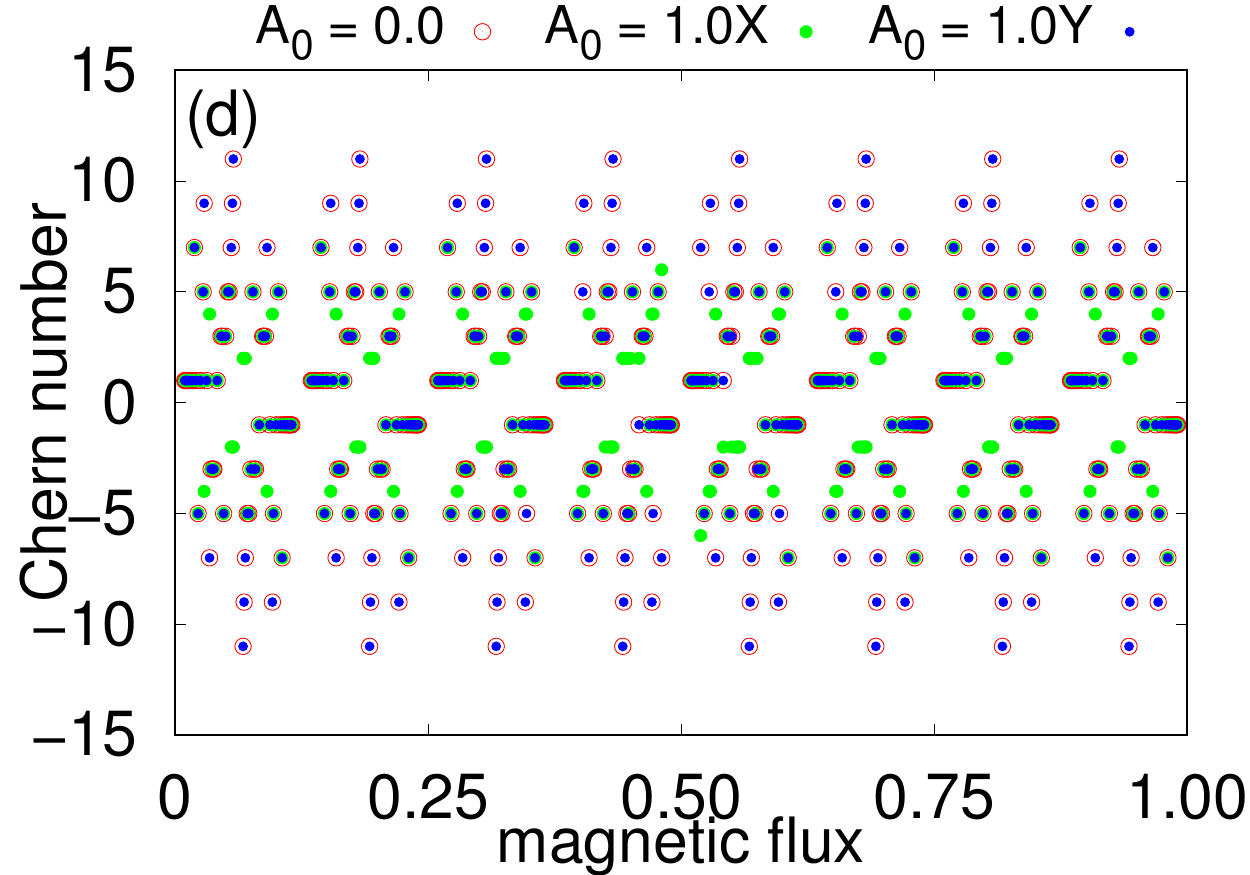}  %
\caption{(Color online.) The ground state Chern number of the Floquet Hofstader butterfly spectrum for the kagome lattice. Laser frequency is fixed at an off-resonance value of $\Omega = 9.0$. (a) circularly polarized light with $A_0 =0.0, 1.0, 2.0$.
(b) linearly polarized light along $x$ ($\alpha=0.0$), $A_0 = 0.0, 1.0, 2.0$. (c) linearly polarized light along $y$  ($\alpha = \pi/2$), $A_0 = 0.0, 1.0, 2.0$. (d) linearly polarized light $A_0 = 1.0$ along $x$ and $y$ as a comparison. The calculations are done with 9 Floquet copies. The magnetic flux is defined as $\phi = p/8q$ with $p$ ranging from 1 to $8q-1$ and $q \leq 13$. 
}
\label{fig:kagome-fig4}
\end{figure*}
\subsection{Equilibrium Hamiltonian with static magnetic field}
In the presence of a vector potential the hopping
parameter $t_h$ gets modified by the Peierls phase,
\begin{align}
   t_h \mapsto t_h e^{i\theta_{ij}}\,,
\label{PeierlsPhase}
\end{align}
where the phase is the integral over the vector potential along the
hopping path.
\begin{align}
     \theta_{ij} = - \frac{e}{\hbar c} \int_{r_i}^{r_j} \vec{A}(\vec{r}) \cdot d\vec{r} = - \frac{2\pi}{\phi_0} \int_{r_i}^{r_j} \vec{A}(\vec{r}) \cdot d\vec{r}
\end{align}
where $\phi_0 \equiv hc/e = 1$ is the magnetic-flux quantum. We define $\phi = B S$ as the magnetic flux through the smallest triangle in one unit cell where $S = a^2/2$.  
The Landau gauge $\vec{A}(\vec{r}) = (0,Bx,0)$ is adopted and the corresponding magnetic unit cell (enlarged parallelogram) is shown in Fig.\ref{fig:magnetic-unitcell}. The hopping phases are 
\begin{align}
\theta_1  = 8m\phi \times (2\pi), \quad
\theta_2 = \phi \times (2\pi)&,\quad\nonumber\\
\theta_3 = (8m-1)\phi \times (2\pi),&
\label{peierls}
\end{align}
and $\theta = 0$ for the hopping along other bonds.
To satisfy the periodic boundary conditions, the uniform-flux strength for the kagome lattice is given by 
\begin{align}
\phi = p/(8q).
\label{eq:flux}
\end{align}

The explicit form of the Hamiltonian is,
\begin{align}
H = H^{(1)} + H^{(2)} + H^{(3)}. 
\end{align}
To recover the translational symmetry of the lattice, we enlarge the unit cell along the translational vector $\vec a_1$ of the original unit cell  by a factor of $q$ and rewrite the position of each unit cell as,
\begin{align}
\tilde{R}(m,n) = m \vec{a}_1 \times q + n \vec{a}_2.
\label{eq:vector-magnetic}
\end{align}
The relation between original, Eq.\eqref{eq:vector-original}, and the enlarged, Eq.\eqref{eq:vector-magnetic}, unit cell vectors is,
\begin{align}
R(m,n) = \tilde{R}(m',n) + (l-1) \vec{a}_1,
\end{align}
with $m' = (m-1)/q + 1$ and $l = {\rm mod}(m-1,q) + 1$.
The Hamiltonian can be rewritten in the enlarged unit cell as,
\begin{align}
H^{(1)} =& -\tsh \sum_{mn}\sum_{l=1}^{q} c^{\dagger}_{(m,n),l} a_{(m,n),l}^{} \nonumber\\
&-\tsh \sum_{mn}\sum_{l=1}^{q} c^{\dagger}_{(m,n),l} a_{(m,n+1),l}^{}, \nonumber\\ 
\end{align}
\begin{align}
 H^{(2)} =& -\tsh \sum_{mn}\sum_{l=1}^{q} c^{\dagger}_{(m,n),l} b_{(m,n),l}^{} \nonumber\\
 &-\tsh \sum_{mn}\sum_{l=1}^{1} c^{\dagger}_{(m,n),1} b_{(m-1,n+1),q}^{} \nonumber\\ 
& -\tsh \sum_{mn}\sum_{l=2}^{q} c^{\dagger}_{(m,n),l} b_{(m,n+1),l-1}^{}, \nonumber\\
\end{align}
\begin{align}
H^{(3)} =& -\tsh \sum_{mn}\sum_{l=1}^{q} b^{\dagger}_{(m,n),l} a_{(m,n),l}^{} \nonumber\\
&-\tsh \sum_{mn}\sum_{l=1}^{q-1} b^{\dagger}_{(m,n),l}a_{(m,n),l+1}^{} \nonumber\\ 
&  -\tsh \sum_{mn}\sum_{l=q}^{q} b^{\dagger}_{(m,n),q} a_{(m+1,n),1}^{}.
\end{align}
Consider the magnetic phase (a gauge choice),
\begin{align}
H^{(1)} =& -\tsh \sum_{mn}\sum_{l=1}^{q} c^{\dagger}_{(m,n),l} a_{(m,n),l}^{}  \nonumber\\
&-\tsh \sum_{mn}\sum_{l=1}^{q} c^{\dagger}_{(m,n),l} a_{(m,n+1),l}^{} e^{-i2\pi (8l)\phi},\nonumber\\ 
\end{align}
\begin{align}
 H^{(2)} =& -\tsh \sum_{mn}\sum_{l=1}^{q} c^{\dagger}_{(m,n),l} b_{(m,n),l}^{} e^{+i2\pi \phi} \nonumber\\
 &-\tsh \sum_{mn}\sum_{l=1}^{1} c^{\dagger}_{(m,n),1} b_{(m-1,n+1),q}^{}{e^{-i2\pi (8l-1)\phi}} \nonumber\\ 
& -\tsh \sum_{mn}\sum_{l=2}^{q} c^{\dagger}_{(m,n),l} b_{(m,n+1),l-1}^{}{e^{-i2\pi (8l-1)\phi}},\nonumber\\
\end{align}
\begin{align}
H^{(3)} =& -\tsh \sum_{mn}\sum_{l=1}^{q} b^{\dagger}_{(m,n),l} a_{(m,n),l}^{} \nonumber\\
&-\tsh \sum_{mn}\sum_{l=1}^{q-1} b^{\dagger}_{(m,n),l} a_{(m,n),l+1}^{} \nonumber\\ 
&  -\tsh \sum_{mn}\sum_{l=q}^{q} b^{\dagger}_{(m,n),q} a_{(m+1,n),1}^{}.
\end{align}
After Fourier transformation, 
\begin{align}
H^{(1)} =& -\tsh \sum_{\bfk}\sum_{l=1}^{q} c^{\dagger}_{\bfk,l} a_{\bfk,l}^{} \nonumber\\
&-\tsh \sum_{\bfk}\sum_{l=1}^{q} c^{\dagger}_{\bfk,l} a_{\bfk,l}^{} e^{+i\bfk\cdot \tilde{R}(0,1)} e^{-i2\pi (8l)\phi},\nonumber\\
\end{align}
\begin{align}
H^{(2)} =& -\tsh \sum_{\bfk}\sum_{l=1}^{q} c^{\dagger}_{\bfk,l} b_{\bfk,l}^{} e^{+i2\pi \phi} \nonumber\\
&-\tsh \sum_{\bfk}\sum_{l=1}^{1} c^{\dagger}_{\bfk,1} b_{\bfk,q}^{} e^{+i\bfk\cdot \tilde{R}(-1,1)} {e^{-i2\pi (8l-1)\phi}} \nonumber\\ 
& -\tsh \sum_{\bfk}\sum_{l=2}^{q} c^{\dagger}_{\bfk,l} b_{\bfk,l-1}^{} e^{+i\bfk\cdot \tilde{R}(0,1)} {e^{-i2\pi (8l-1)\phi}},  \nonumber\\
\end{align}
\begin{align}
H^{(3)} =& -\tsh \sum_{\bfk}\sum_{l=1}^{q} b^{\dagger}_{\bfk,l} a_{\bfk,l}^{} \nonumber\\
&-\tsh \sum_{\bfk}\sum_{l=1}^{q-1} b^{\dagger}_{\bfk,l} a_{\bfk,l+1}^{} \nonumber\\ 
&  -\tsh \sum_{\bfk}\sum_{l=q}^{q} b^{\dagger}_{\bfk,q} a_{\bfk,1}^{} e^{+i\bfk\cdot \tilde{R}(1,0)},
\label{eq:Hmag}
\end{align}
where 
\begin{align}
\tilde{R}(+0, 1) &= \bfk\cdot\vec{a}_2,\\
\tilde{R}(-1, 1) &= -\bfk\cdot q\vec{a}_1 + \bfk\cdot\vec{a}_2,\\
\tilde{R}(+1, 0) &= \bfk\cdot q\vec{a}_1.
\end{align}
The energy spectrum is obtained by numerically diagonalizing the $3q\times 3q$ Hamiltonian matrix for each wave vector $\bfk$.
\subsection{Time-dependent Hamiltonian with laser and static magnetic field}
When the system is exposed to laser light, the Hamiltonian for the kagome lattice is rewritten as,
\begin{align}
H^{(1)} =& -\tsh \sum_{\bfk}\sum_{l=1}^{q} c^{\dagger}_{\bfk,l} a_{\bfk,l}^{} e^{i \bfA(t) \cdot \delta_2} \nonumber\\
&-\tsh \sum_{mn}\sum_{l=1}^{q} c^{\dagger}_{\bfk,l} a_{\bfk,l}^{} e^{+i\bfk\cdot R(0,1)} e^{-i2\pi (8l)\phi} e^{-i \bfA(t) \cdot \delta_2}, \nonumber\\
\end{align}
\begin{align}
H^{(2)} =& -\tsh \sum_{\bfk}\sum_{l=1}^{q} c^{\dagger}_{\bfk,l} b_{\bfk,l}^{} e^{+i2\pi \phi} e^{i \bfA(t) \cdot \delta_3}  \nonumber\\
&-\tsh \sum_{\bfk}\sum_{l=1}^{1} c^{\dagger}_{\bfk,1} b_{\bfk,q}^{} e^{+i\bfk\cdot R(-1,1)} {e^{-i2\pi (8l-1)\phi}}  e^{-i \bfA(t) \cdot \delta_3} \nonumber\\ 
& -\tsh \sum_{\bfk}\sum_{l=2}^{q} c^{\dagger}_{\bfk,l} b_{\bfk,l-1}^{} e^{+i\bfk\cdot R(0,1)} {e^{-i2\pi (8l-1)\phi}} e^{-i \bfA(t) \cdot \delta_3},   \nonumber\\
\end{align}
\begin{align}
H^{(3)} =& -\tsh \sum_{\bfk}\sum_{l=1}^{q} b^{\dagger}_{\bfk,l} a_{\bfk,l}^{} e^{+i \bfA(t) \cdot \delta_1} \nonumber\\
&-\tsh \sum_{\bfk}\sum_{l=1}^{q-1} b^{\dagger}_{\bfk,l} a_{\bfk,l+1}^{} e^{-i \bfA(t) \cdot \delta_1}  \nonumber\\ 
&  -\tsh \sum_{\bfk}\sum_{l=q}^{q} b^{\dagger}_{\bfk,q} a_{\bfk,1}^{} e^{+i\bfk\cdot R(1,0)} e^{-i \bfA(t) \cdot \delta_1}. 
\end{align}

The time-dependent Hamiltonian can be solved numerically within the framework of Floquet theory. The standard process for generating the time-independent Floquet Hamiltonian can be found in Ref.~[\onlinecite{Rudner:prx13}].
One sub-block of the Floquet Hamiltonian is given by,
\begin{align}
H_{n,m} = \frac{1}{T} \int_{0}^{T} dt \exp\{i(n-m)\Omega t\} H(t), 
\end{align}
where $n,m$ are the Floquet replica numbers.
We need to calculate the expression with the general form,
\begin{align}
      f_{nm} = \frac{1}{T} \int_0^T dt e^{-i(n-m)\Omega t}  \exp[-i \bfA(t)\cdot{\bf d}].
\end{align}
Here we use ${\bf d}={\bf R}_j-{\bf R}_i$, and define $d^x /|{\bf d}| = \cos\theta$, $d^y/|{\bf d}|= \sin\theta$. For nearest-neighbor hopping terms, $|{\bf d}|=1$, $\theta = 0, 2\pi/3,  -2\pi/3$.

Substituting the vector potential into the above equation,
\begin{align}
          &\frac{1}{T} \int_0^T dt e^{-i(n-m)\Omega t}  \exp[-i A_0 (d^x \sin\Omega t + d^y \cos\Omega t)] \nonumber\\
      =&\mathcal{J}_{m-n}(A_0 |{\bf d}|) \exp[i(n-m)\theta],
\end{align}
where $\mathcal{J}_{n}(x)$ is a Bessel function of the first kind.

Substituting the vector potential for linearly polarized light $\bfA(t)=A_0\sin(\Omega t) (\cos\alpha, \sin\alpha)$ into the above equation gives,
\begin{align}
          &\frac{1}{T} \int_0^T dt e^{-i(n-m)\Omega t}  \exp[-i A_0 (d^x cos\alpha + d^y \sin\alpha) \sin(\Omega t)] \nonumber\\
      = &\mathcal{J}_{m-n}(A_0 |{\bf d}| \cos(\theta-\alpha)),
\end{align}
which describes the renormalization of the hopping parameters along the different directions at lowest order. 

\section{Hofstadter Butterfly on the kagome lattice}
\label{sec:butterfly}
In the equilibrium case (without laser light), we calculate the Hofstadter butterfly on the kagome lattice by setting $q=199$ [Eq.\eqref{eq:flux}] in Fig.\ref{fig:kagome-fig1}(a). 
The energy spectrum versus static magnetic flux (the Hofstadter butterfly) is calculated by diagonalizing the Hamiltonian, Eq.\eqref{eq:Hmag}, at the $\Gamma$ point ($k_x = 0, k_y = 0$) for varying flux $\phi=p/8q$. 
There exist $3q$ magnetic mini-bands in one unit cell. 
Previous studies on the square and honeycomb lattices with an isotropic hopping integral describe rich symmetries in the Hofstadter butterfly.  For example, a reflection symmetry about flux $\phi=1/2$ and a reflection symmetry about energy $E=0$. 

For the kagome lattice, we observe the reflection symmetry about the energy axis $E=0$ is lacking because the particle-hole symmetry is broken. The reflection symmetry about the flux $\phi=1/2$ is observed $E(\phi) = E(1-\phi)$, where we used $E(\phi)$ to denote the energy spectrum of Hamiltonian [Eq.\eqref{eq:Hmag}] with magnetic flux $\phi$. This is because the reflection symmetry about $\phi=1/2$ is preserved by time-reversal symmetry and changing the lattice will not break the symmetry.\cite{Osadchy:jmp01} The time reversal operation of $H(\phi)$ is $H(-\phi)$, and we can see $H(1-\phi) = H(-\phi)$ from Eq.\eqref{eq:Hmag}. So we have
\begin{align}
T H(\phi) T^{-1} = H (-\phi) = H(1-\phi),
\end{align}
where $T$ is the time-reversal operator, which is anti-unitary. 
Since the two operators which are time-reversal partners will have the same eigenvalues, the symmetry about $\phi=1/2$ is explained.  We further observe the central symmetry about $\phi=1/4$, which is $E(\phi) = - E(1/2 - \phi)$. We do not have a simple physical picture to explain this symmetry property in the spectrum.

In equilibrium studies, Hofstadter's butterfly is often plotted over the flux region $0<\phi<1/2$, because the remaining part, $1/2<\phi<1$, is just the mirror image of the previous part.\cite{LiJ:jpcm11} In the driven system, by contrast, the external drive can break the time reversal symmetry. For example, circularly polarized light breaks time-reversal symmetry, while linearly polarized light preserves it. We show the full Hofstadter butterfly in the magnetic flux region $0<\phi<1$. 

The effects of off-resonant ($\hbar\Omega = 9.0$) circularly polarized laser light are shown in Fig.\ref{fig:kagome-fig1}(b), (c), (d) for laser amplitudes $A_0 = 1.0, 2.0, 3.5$, respectively.
As the laser intensity increases, the bandwidth first decreases. Afterwards, the bandwidth then increases, but the bands are inverted (there is a sign change in the effective hopping parameter). 
This behavior can be understood using the Floquet-Magnus expansion in the high frequency regime, 
\begin{align}
H_{\text{eff}} = H_0 + \frac{1}{\hbar\Omega} [H_1, H_{-1}] + \cdots
\end{align}
where $H_n = \frac{1}{T}\int_{0}^{T} e^{-i n\Omega t} H(t) dt$.
In the theoretical infinite frequency limit, the Floquet-Bloch band is the original one scaled by a zeroth order Bessel function (zeroth order term in the Floquet-Magnus expansion). 
The Floquet butterfly spectrum will have decreased bandwidth with increasing laser intensity until $A_0 = 2.404$, which is the first zero point of the zeroth-order Bessel function. 
After that point, the band will be inverted with increasing bandwidth up to $A_0=3.8$. Our Floquet butterfly is consistent with the high frequency analysis, except some structural details are different. A systematic analysis needs to include the higher-order terms in the Magnus expansion. 

Let us dive into the details of the effects of circularly polarized light on the Hofstadter butterfly. The reflection symmetry about flux $\phi=1/2$ is broken. This phenomenon is explained qualitatively by, 
\begin{align}
     T H(\phi, \vec{A}_L(t)) T^{-1} =  H(-\phi, \vec{A}_R(t)),
\end{align}
where the time-reversal of left circularly polarized light is its right polarized partner, as indicated by the subscript on the vector potential, $\vec A$. 
From the numerical data, one can see the central symmetry about $\phi=1/4$ is preserved for circularly polarized light. 

In Fig.\ref{fig:kagome-fig2}, we plot the energy spectrum as a function of magnetic flux for light linearly polarized along the $x$ Fig.\ref{fig:kagome-fig2}(a,c) or $y$ Fig.\ref{fig:kagome-fig2}(b,d) direction, and for laser intensity $A_0=1.0$ or $A_0 =2.0$. From the numerical data shown, we conclude the energy spectrum is polarization direction dependent. 

In Fig.\ref{fig:kagome-fig3}, we plot the energy spectrum as a function of magnetic flux for laser frequency in the on-resonance regime ($\hbar\Omega = 4.0$). The data plotted with black dots are the spectrum of the central (in energy) Floquet copy. The data plotted with red dots are the spectrum of the upper and lower Floquet copies. In Fig.\ref{fig:kagome-fig3} (a) and (c), the spectrum as a function of magnetic flux for circularly polarized light is plotted for laser intensity $A_0 = 1.0$ and $A_0 = 2.0$ respectively. The reflection symmetry is broken about $\phi = 1/2$ , while the inversion symmetry about $\phi=1/4$ is preserved, as we have observed in the off-resonance laser frequency region.  

As a comparison, the spectrum as a function of magnetic flux for linearly polarized light is plotted for laser intensity $A_0 = 1.0$ and $A_0 = 2.0$ in Fig.\ref{fig:kagome-fig3} (b) and (d), respectively.  For linearly polarized light, both the reflection symmetry about $\phi = 1/2$ and the inversion symmetry about $\phi=1/4$ are preserved. 


\section{Spin Chern number for the Kagome lattice}
\label{sec:chern}
Following Ref.[\onlinecite{Goldman:jpb09}] and Ref.[\onlinecite{Wackerl:arXiv18}], we calculate the Chern number of the ground state of the Hofstadter butterfly, where the ``ground state'' in Floquet-Bloch band structure shall be understand as the lowest energy band of the central Floquet copy. 
The Chern number data are calculated using Fukui's method.\cite{Fukui:jpsj05}
To avoid the band crossing between different Floquet copies, we fix the laser frequency to be in the off-resonance region. 
In Fig.\ref{fig:kagome-fig4}, we plot the Chern number for the ``ground state'' of the Hofstadter butterfly with laser frequency fixed at $\hbar\Omega = 9.0$.  In these Chern number plots, we also plot the data with vanishing laser intensity as a reference point. First, consider the reference point at $A_0=0$. Because $H(\phi)$ is the time-reversal partner of $H(1-\phi)$, we have the symmetry structure of the Chern numbers as $C(\phi) = - C(1-\phi)$. 

In Fig.\ref{fig:kagome-fig4}(a) the data is shown for circularly polarized light with parameter $A_0 = 1.0$ and $A_0= 2.0$. 
From the numerical data, we realize the Chern numbers calculated with laser intensity $A_0 = 1.0$ are the same as those for vanishing laser intensity, which means that while the band structure is deformed under the circularly polarized light, the Chern numbers still preserve the properties of ``time-reversal symmetry" about $\phi=1/2$ (as discussed earlier in the manuscript) for low laser intensity. Further increasing the laser intensity to $A_0=2.0$ will show some difference; the Chern numbers differ somewhat from the reference points, especially at larger magnetic fluxes. 

We now consider the effect of linearly polarized light.  Fig.\ref{fig:kagome-fig4}(b) and (c) show the data for linearly polarized light with parameter $A_0 = 1.0, 2.0$, and polarization direction along the $x$ (b) and $y$ (c) directions, respectively. In contrast to the circularly polarized light, the linearly polarized light will preserve time-reversal symmetry (in the absence of the static magnetic flux on the lattice): we have $C(\phi) = - C(1-\phi)$.  When the polarization direction is along the $x$-axis, we find the Chern numbers for different laser intensities are different.  On the other hand, if the polarization direction is along the $y$ axis, that data for $A_0 = 1.0$ appears numerically similar, while the data for $A_0=2.0$ differs around $\phi=1/2$. 
As pointed out in Ref.[\onlinecite{Wackerl:arXiv18}], for circularly polarized light the ``ground state" is uniquely defined, whereas for linearly polarized light it is not uniquely defined for all flux values. Here we find a band crossing with the ``ground state'' occurs at magnetic flux $p/8q = 39/88, 45/104, 46/104, 49/104, 50/104$. For clarity, Fig.\ref{fig:kagome-fig4}(d) compares Chern numbers for linearly polarized light along the $x$ and $y$ directions for fixed laser intensity. 

\begin{figure*}[t]
\includegraphics[width=0.99\linewidth, angle=0]{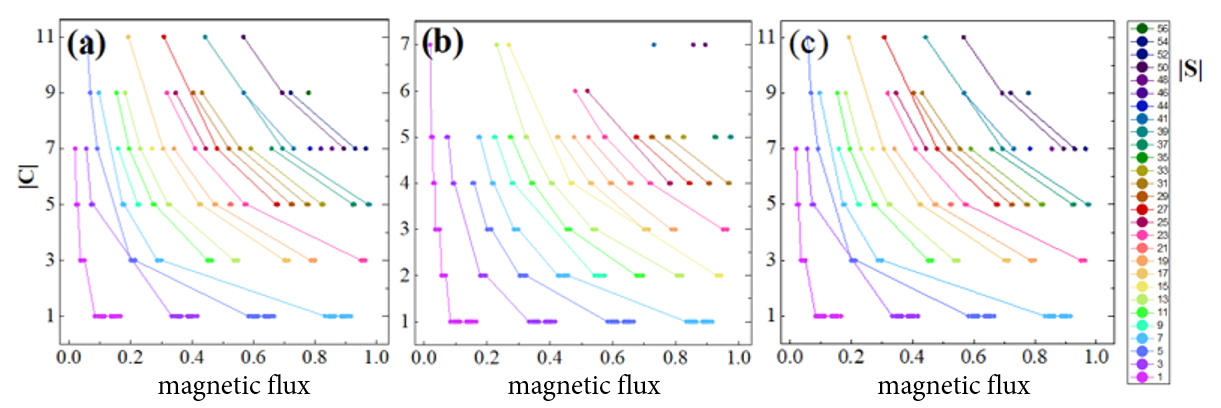}  %
\caption{(Color online.) Chern number $|C_n|$ as a function of the magnetic flux $\phi$ for the kagome lattice exposed to circularly (a) and linearly (b,c) polarized lasers with the laser frequency fixed to be in the off-resonant regime, $\Omega = 9.0$. (a) circularly polarized light $A_0 = 1.0$.
(b) linearly polarized light $A_0 = 1.0$, $\alpha = 0$. (c) linearly polarized light $A_0 = 1.0$, $\alpha = \pi /2$. The calculations are done with 9 Floquet copies. The magnetic flux is defined as $\phi = p/8q$ with $p$ ranging from 1 to $8q-1$ and $q \leq 13$.}
\label{fig:diophantine-kagome}
\end{figure*}

Since the magnetic-translation symmetry is preserved as the system is exposed to an external laser, the topological invariant must satisfy the Diophantine equation, \cite{Lababidi:prl14,Dana:jpc85,Kooi:arXiv18}
\begin{align}
s = \frac{1}{q} + \frac{p}{q} C,
\end{align}
for flux $\phi = p/8q$, where $C$ is the topological invariant and $s$ is an integer. We have verified that our calculated Chern numbers satisfy the Diophantine equation; a representative subset is displayed in the graphs in Fig.\ref{fig:diophantine-kagome}.
Following the reference [\onlinecite{Goldman:jpb09}], we connect all the points ($\phi=p/8q$, $|C|$) that are associated with the same number $|s|$ with a colored line. As $|s|$ increases, the color changes progressively from fuchsia to teal.

\section{Hofstadter Butterfly on the Triangular Lattice}
\label{sec:triangular}
Because the triangular lattice has the same Bravais lattice as the kagome lattice, and there is only one atom in each unit cell (in the absence of a static magnetic field), the Hofstadter butterfly on the triangular lattice can have features similar to the kagome lattice butterfly.\cite{LiJ:jpcm11}  Here we have studied the Floquet Hofstadter butterfly on the triangular lattice, and the results are indeed qualitatively similar to those for the kagome lattice. 

\begin{figure}[h]
\includegraphics[width=0.99\linewidth, angle=0]{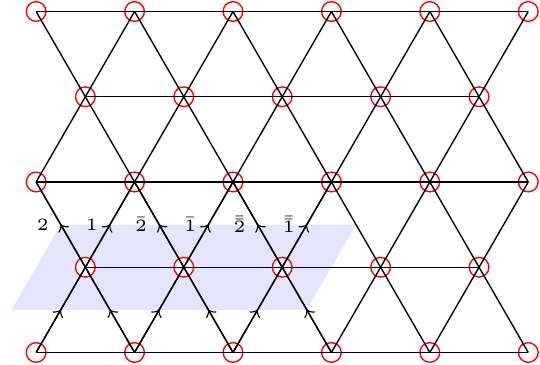}  %
\caption{(Color online.) In the case without magnetic field, the triangular lattice with nearest neighbor vectors $\vec{\delta}_1 = (1,0)a$, $\vec{\delta}_2 = (1/2, \sqrt{3}/2)a$, $\vec{\delta}_3 = (-1/2, \sqrt{3}/2)a$, where $a$ the nearest neighbor distance in triangular lattice, is plotted. The translational vectors are $\vec{a}_1 = \vec{\delta}_1$ and $\vec{a}_2 = \vec{\delta}_2$. The reciprocal lattice vectors are $\vec{b}_1 = (1, -1/\sqrt{3})2\pi/a$ and $\vec{b}_2 = (0, 2/\sqrt{3})2\pi/a$.
When the system is exposed to a perpendicular magnetic field, the magnetic unit cell must be enlarged (to recover the translational symmetry) depending on the value of the magnetic flux, $\phi$. For example, the magnetic cell is the blue area (shaded parallelogram) for magnetic flux $\phi/\phi_0 = 1/2q$, with $q=3$ ($\phi_0$ is defined as magnetic flux quantum).}
\label{fig:magnetic-unitcell-tri}
\end{figure}

\begin{figure*}[t]
\includegraphics[width=0.49\linewidth, angle=0]{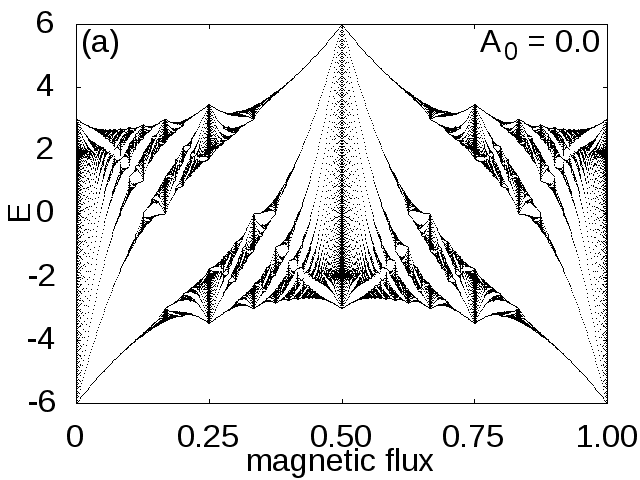}  %
\includegraphics[width=0.49\linewidth, angle=0]{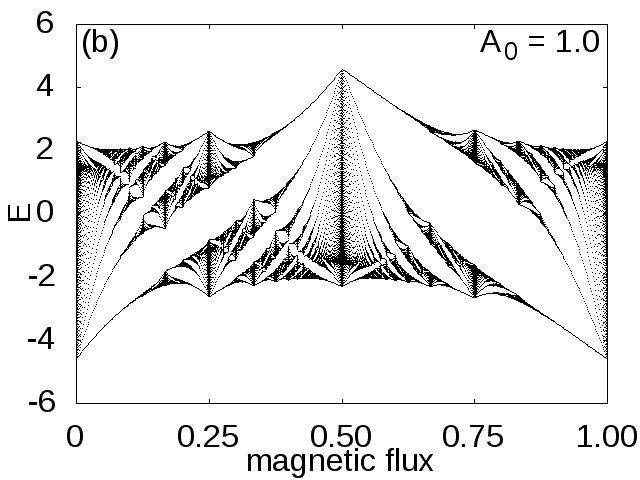}  %
\includegraphics[width=0.49\linewidth, angle=0]{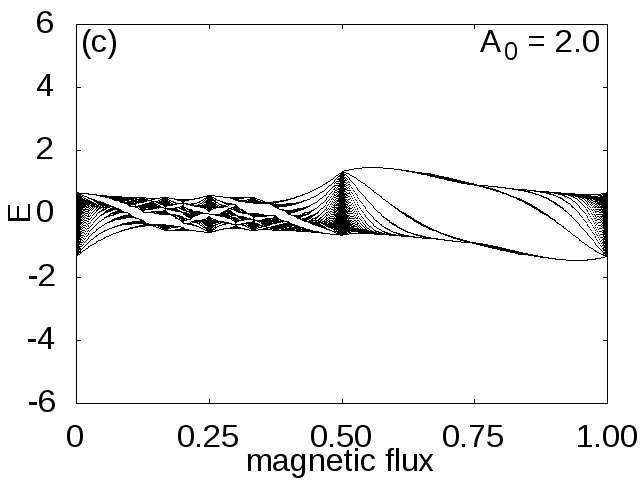}  %
\includegraphics[width=0.49\linewidth, angle=0]{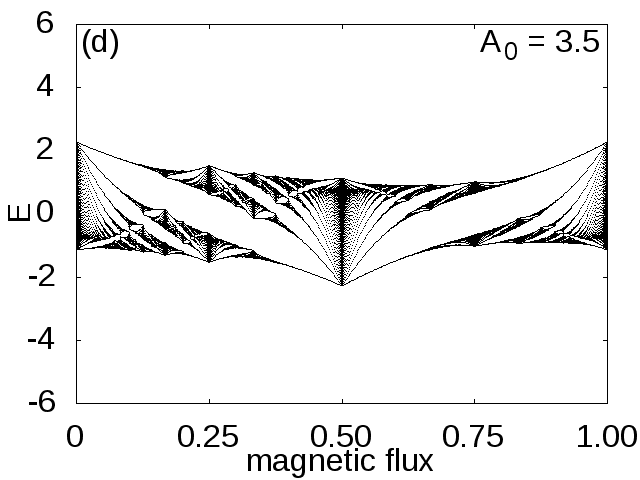}  %
\caption{(Color online.) The Hofstader butterfly on the triangular lattice, deformed by circularly polarized laser light with frequency fixed to be in the off-resonance regime $\Omega = 9.0$. (a) laser intensity $A_0 = 0.0$ (b) laser intensity $A_0 = 1.0$ (c) $A_0 = 2.0$ (d) $A_0 = 3.5$. The calculation is done with 9 Floquet copies. The flux is used as $p/2q$ with $N=299$ and p ranging from 1 to $2N-1$.}
\label{fig:tri-fig1}
\end{figure*}

\begin{figure*}[t]
\includegraphics[width=0.49\linewidth, angle=0]{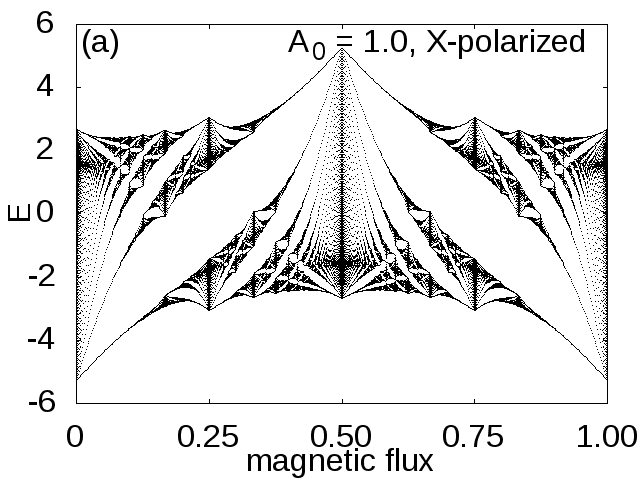}  %
\includegraphics[width=0.49\linewidth, angle=0]{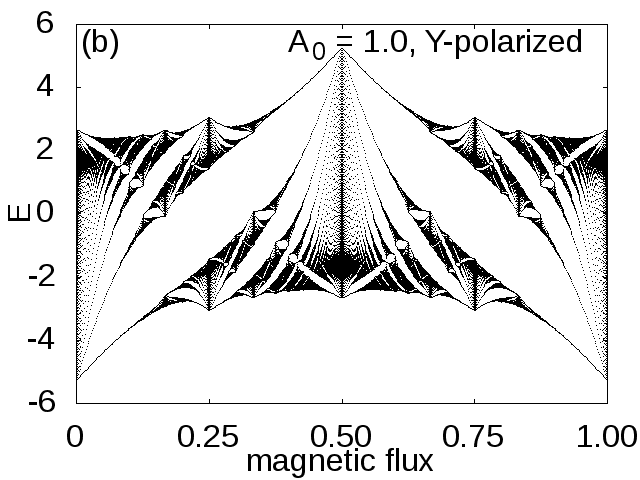}  %
\includegraphics[width=0.49\linewidth, angle=0]{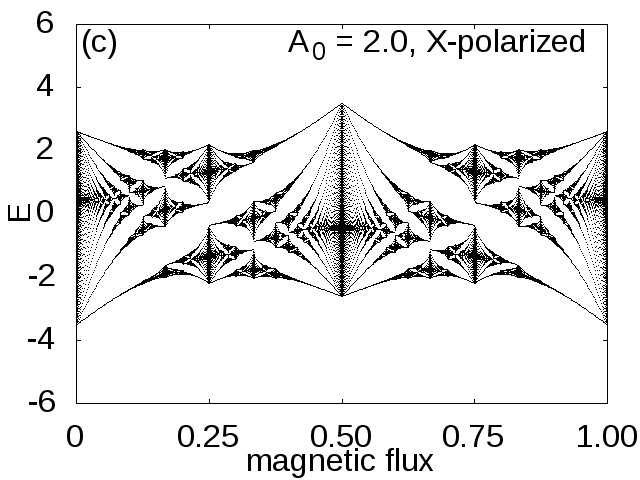}  %
\includegraphics[width=0.49\linewidth, angle=0]{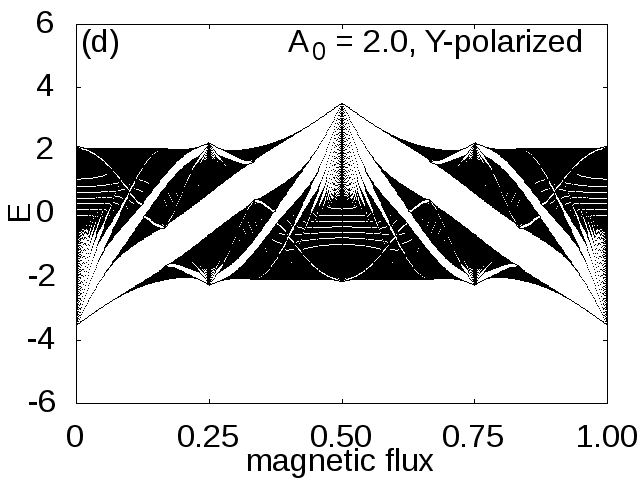}  %
\caption{(Color online.) The Hofstader butterfly for the triangular lattice exposed to a linearly polarized laser with vector potential $A(t) = A_0 \sin(\Omega t)(\cos\alpha, \sin\alpha)$ with laser frequency fixed to be in the  off-resonance regime, $\Omega = 9.0$. (a) $A_0 = 1.0$, $\alpha = 0$.
(b) $A_0 = 1.0$, $\alpha = \pi/2$. (c) $A_0 = 2.0$, $\alpha = 0$. (d) $A_0 = 2.0$, $\alpha = \pi/2$. The calculation is done with 9 Floquet copies. The flux is used as $p/2q$ with $N=299$ and p ranging from 1 to $2q-1$. 
}
\label{fig:tri-fig2}
\end{figure*}

\begin{figure*}[t]
\includegraphics[width=0.49\linewidth, angle=0]{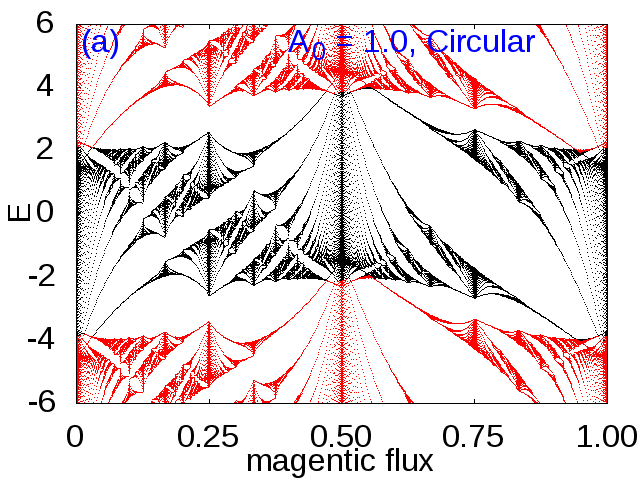}  %
\includegraphics[width=0.49\linewidth, angle=0]{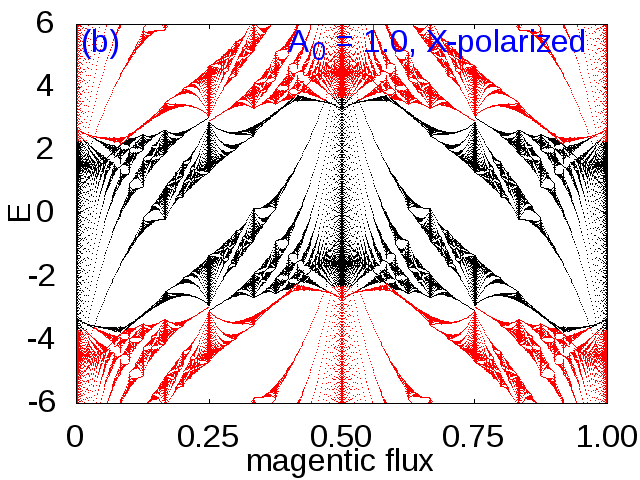}  %
\includegraphics[width=0.49\linewidth, angle=0]{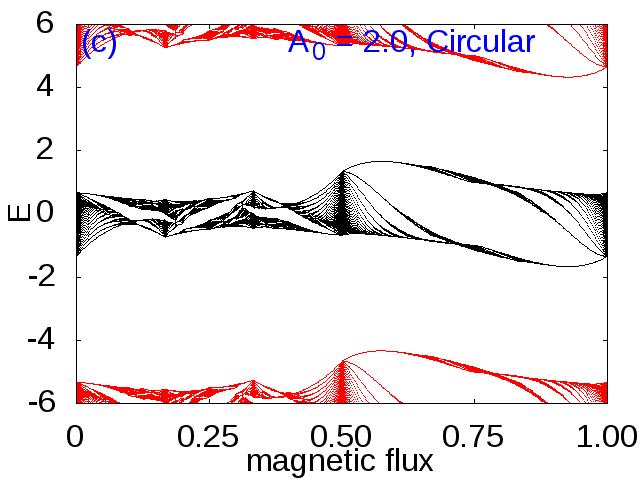}  %
\includegraphics[width=0.49\linewidth, angle=0]{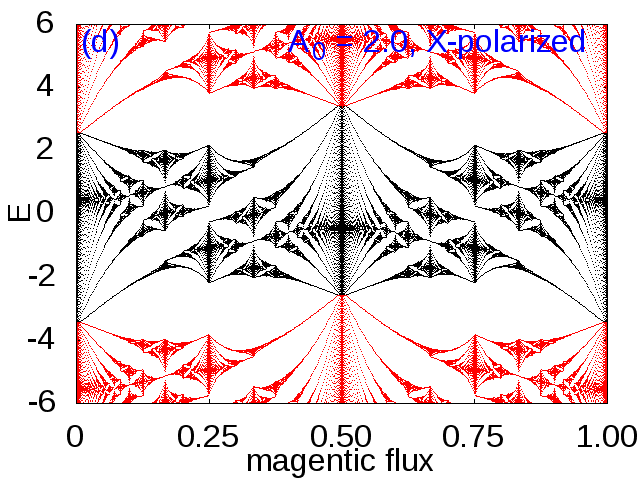}  %
\caption{(Color online.) The Hofstader butterfly for the triangular lattice exposed to circularly (a,c) and linearly (b,d) polarized laser with laser frequency fixed to be in the on-resonance regime, $\Omega = 4.0$. (a) circularly polarized light $A_0 = 1.0$.
(b) linear polarized light $A_0 = 1.0$, $\alpha = 0$. (c) circularly polarized light $A_0 = 2.0$. (d) linear polarized light $A_0 = 2.0$, $\alpha = 0$. The red points are data from upper and lower Floquet copies. The calculation is done with 9 Floquet copies. The flux used is $p/2q$ with $q=299$ and p ranging from 1 to $2q-1$. 
}
\label{fig:tri-fig3}
\end{figure*}

\begin{figure*}[t]
\includegraphics[width=0.49\linewidth, angle=0]{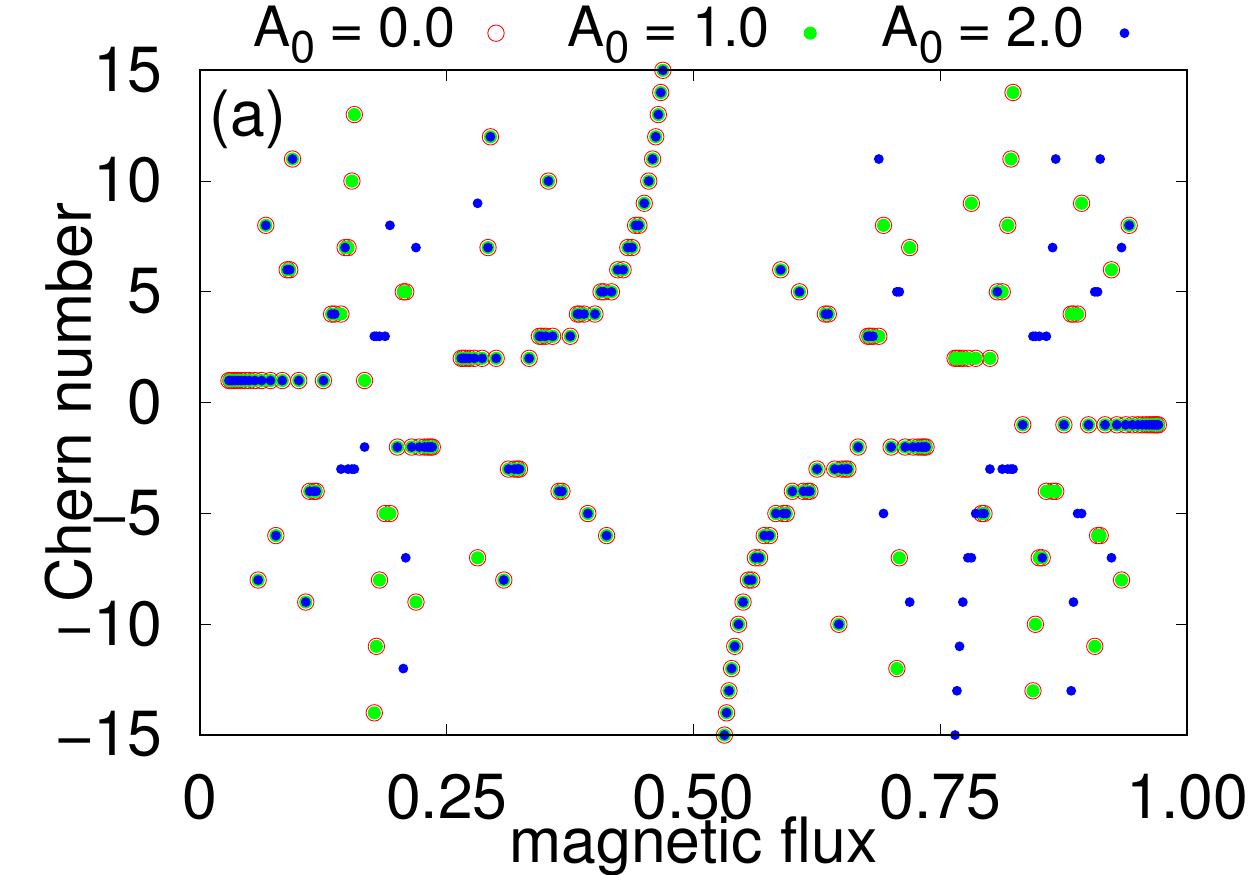}  %
\includegraphics[width=0.49\linewidth, angle=0]{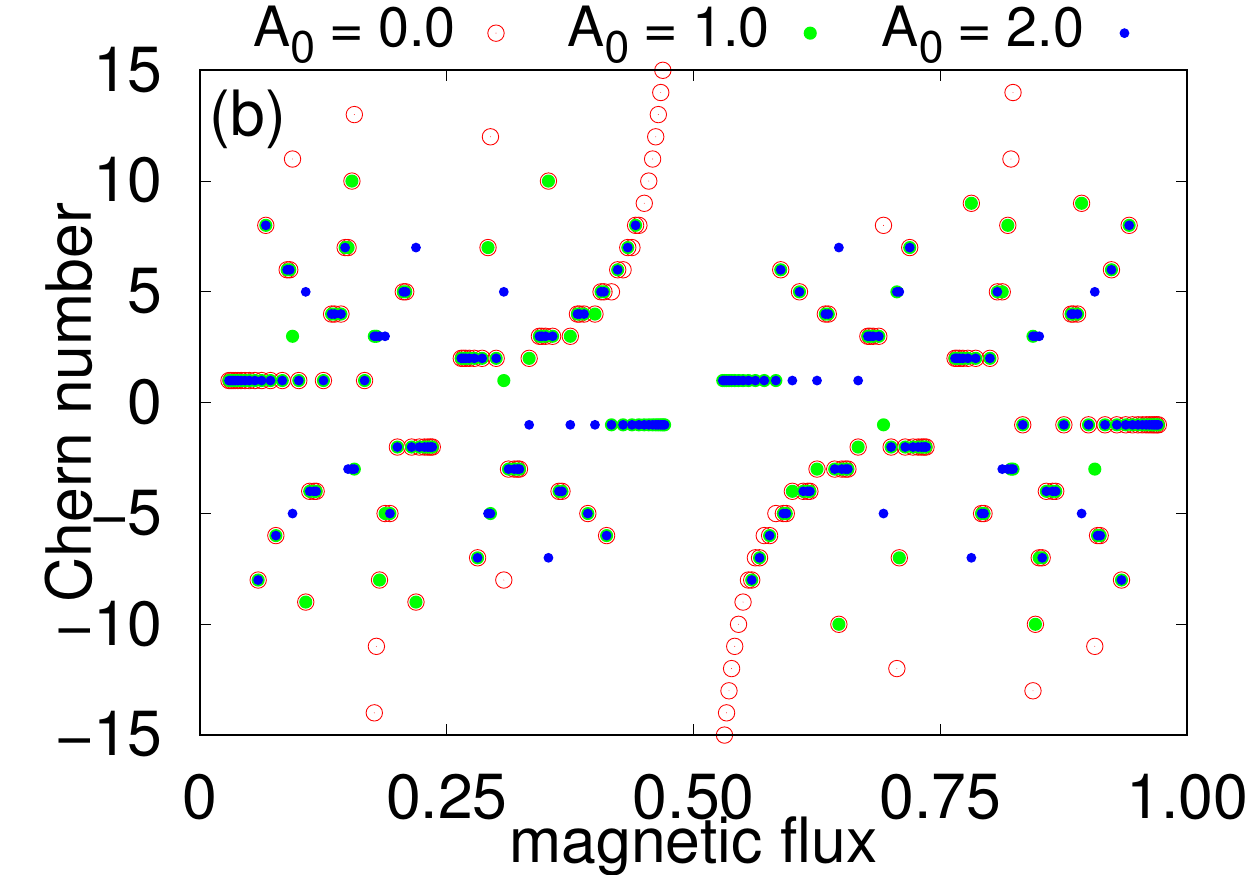}  %
\includegraphics[width=0.49\linewidth, angle=0]{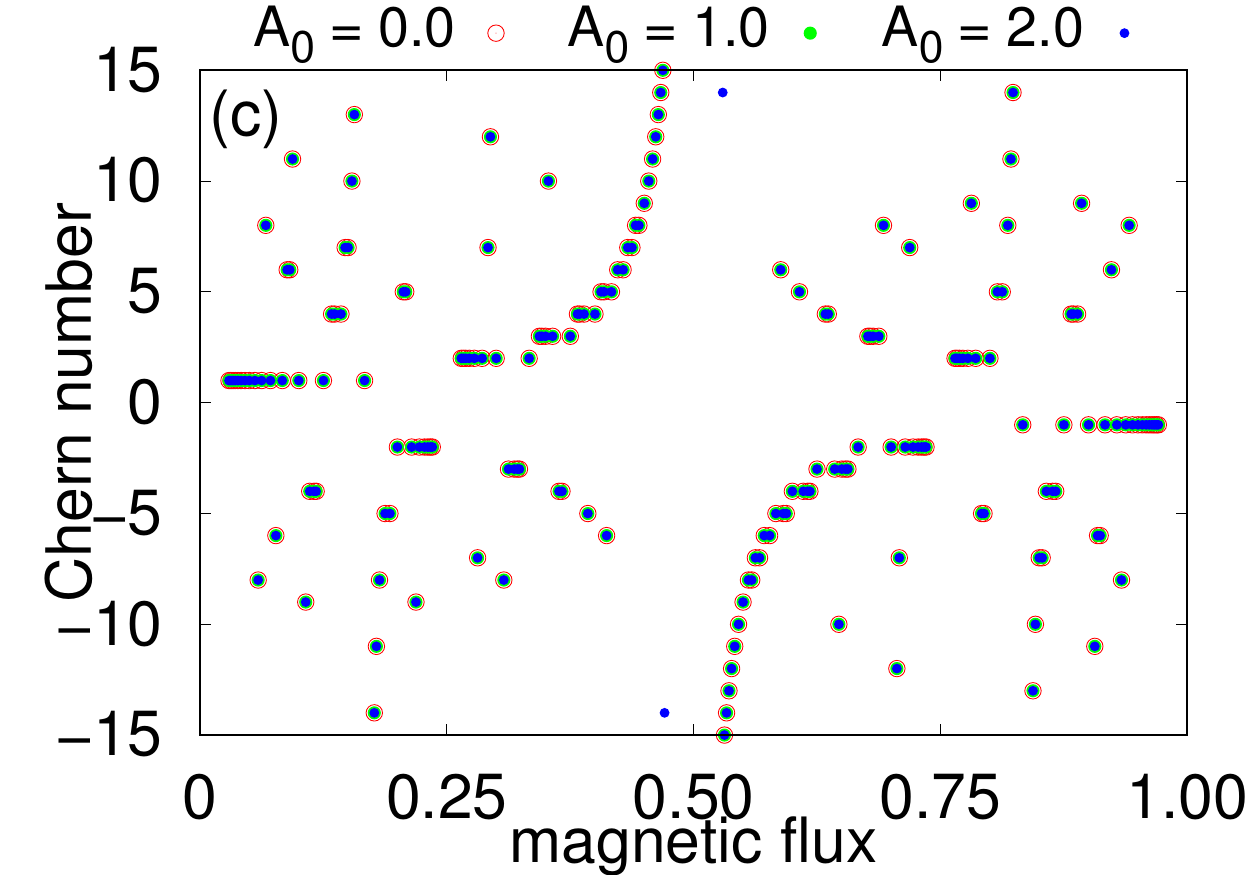}  %
\includegraphics[width=0.49\linewidth, angle=0]{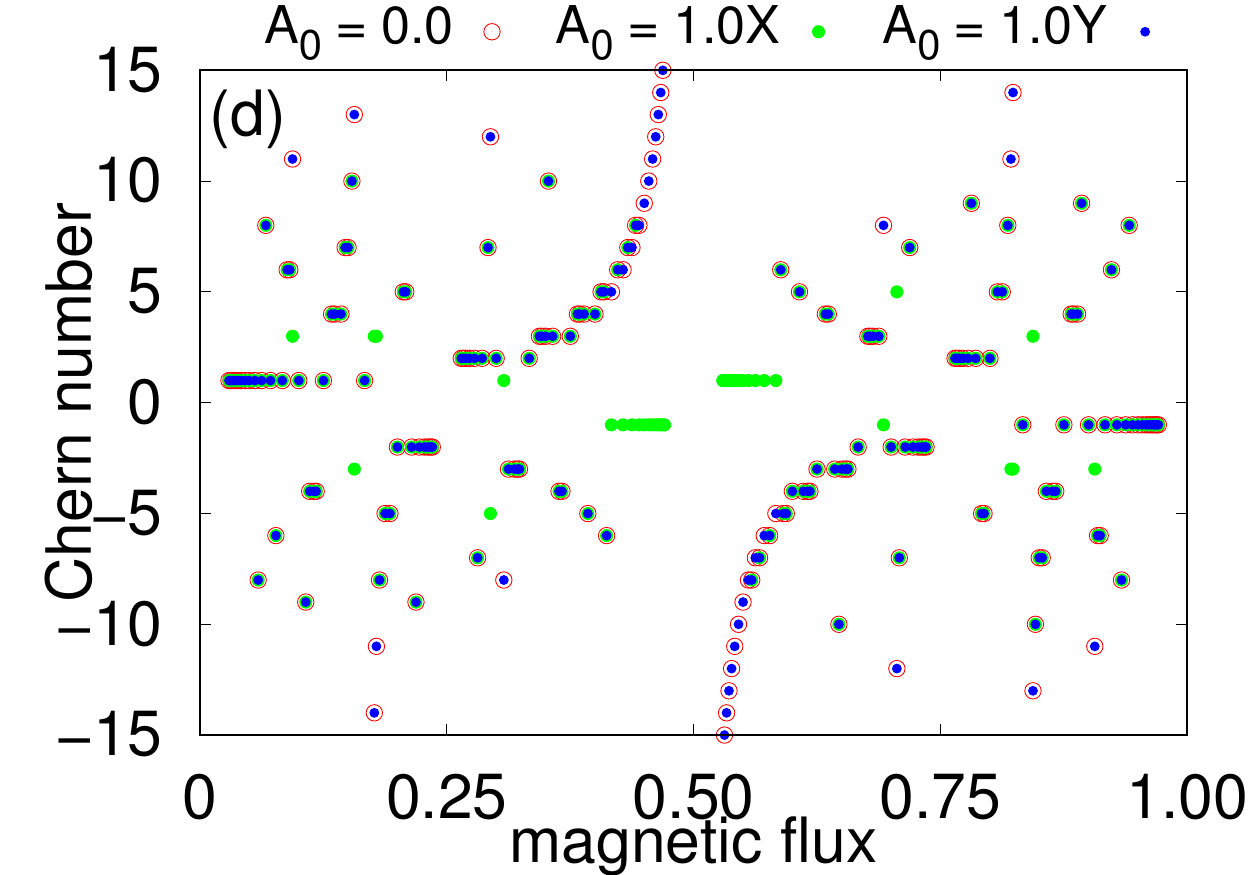}  %
\caption{(Color online) Chern number of the Hofstader butterfly for the triangular lattice exposed to a circularly (a,c) and linearly (b,d) polarized laser with laser frequency fixed to be in the on-resonance regime, $\Omega = 9.0$. (a) circularly polarized light with $A_0 =0, 1, 2$.
(b) linear polarized light along $x$ $\alpha = 0$, $A_0 = 0.0, 1.0, 2.0$. (c) linear polarized light along $y$ $\alpha = \pi/2$, $A_0 = 0.0, 1.0, 2.0$. (d) linear polarized light $A_0 = 1.0$. The calculation is done with 9 Floquet copies. The flux used is $p/2q$ with $q=299$ and p ranging from 1 to $2q-1$. 
}
\label{fig:tri-fig4}
\end{figure*}

In Fig.\ref{fig:magnetic-unitcell-tri}, the triangular lattice and a magnetic unit cell (shaded zone) are shown for magnetic flux $\phi=1/(2\times 3)$. In Fig.\ref{fig:tri-fig1}, the Hofstadter butterfly deformed by off-resonance circularly polarized light is plotted with laser intensities $A_0= 0.0, 1.0, 2.0, 3.5$. The laser frequency is fixed at $\hbar\Omega = 9.0$. 
In Fig.\ref{fig:tri-fig2}, the Hofstadter butterfly deformed by off-resonance linearly polarized light is plotted with laser intensity $A_0= 1.0, 2.0$. The laser frequency is fixed at $\hbar\Omega = 9.0$. The direction of polarization is also considered: we find $x-$ ($y-$) polarized light will have different effects on the spectrum. The same study with different laser frequencies is given in Fig.\ref{fig:tri-fig3},\ref{fig:tri-fig4}. Finally, we study the Chern numbers for the Floquet ``ground state''.



\section{Conclusion}
\label{sec:conclusion}
In this paper, we study the energy spectrum as a function of magnetic flux on the kagome and triangular lattices subjected to a uniform perpendicular magnetic field in the presence of either circularly or linearly polarized light. We find circularly polarized light deforms the Hofstadter butterfly by breaking the reflection symmetry about magnetic flux $\phi = 1/2$, while linearly polarized light preserves that mirror symmetry. This contrasting behavior is explained by the fact that circularly polarized light breaks time-reversal symmetry (in the absence of the static magnetic flux on the lattice), while linearly polarized light preserves the symmetry (in the absence of the static magnetic flux on the lattice).  Further, the inversion symmetry about $\phi=1/4$  is always preserved for both circularly and linearly polarized light. Focusing on linearly polarized light, we find the energy spectrum depends on the polarization direction because the lattice is not isotropic in the $x$ and $y$-directions.

The ground state spin Chern number of the Hofstadter butterfly, where the ``ground state'' in Floquet-Bloch band structure shall be understood as the lowest energy band of the central Floquet copy given a gauge choice, are studied. For circularly polarized light, we conclude that the Chern numbers will coincide with a reference point of vanishing laser intensity for low laser intensity. However, for high laser intensity, the Chern numbers differ. For linearly polarized light, the polarization direction of the light will play a significant role in determining the spin-Chern number.  These behaviors hold for the both the kagome and triangular lattices because the two share the same underlying triangular Bravais lattice.

\acknowledgements 
We acknowledge helpful discussions with Bowen Ma and Xiaoting Zhou. We gratefully acknowledge funding from Army Research Office Grant No. W911NF-14-1-0579, NSF Grant No. DMR-1507621, and NSF Materials Research Science and Engineering Center Grant No. DMR-1720595.  GAF acknowledges support from a Simons Fellowship.
\bibliography{butterfly.bib}

\begin{thebibliography}{39}%
\makeatletter
\providecommand \@ifxundefined [1]{%
 \@ifx{#1\undefined}
}%
\providecommand \@ifnum [1]{%
 \ifnum #1\expandafter \@firstoftwo
 \else \expandafter \@secondoftwo
 \fi
}%
\providecommand \@ifx [1]{%
 \ifx #1\expandafter \@firstoftwo
 \else \expandafter \@secondoftwo
 \fi
}%
\providecommand \natexlab [1]{#1}%
\providecommand \enquote  [1]{``#1''}%
\providecommand \bibnamefont  [1]{#1}%
\providecommand \bibfnamefont [1]{#1}%
\providecommand \citenamefont [1]{#1}%
\providecommand \href@noop [0]{\@secondoftwo}%
\providecommand \href [0]{\begingroup \@sanitize@url \@href}%
\providecommand \@href[1]{\@@startlink{#1}\@@href}%
\providecommand \@@href[1]{\endgroup#1\@@endlink}%
\providecommand \@sanitize@url [0]{\catcode `\\12\catcode `\$12\catcode
  `\&12\catcode `\#12\catcode `\^12\catcode `\_12\catcode `\%12\relax}%
\providecommand \@@startlink[1]{}%
\providecommand \@@endlink[0]{}%
\providecommand \url  [0]{\begingroup\@sanitize@url \@url }%
\providecommand \@url [1]{\endgroup\@href {#1}{\urlprefix }}%
\providecommand \urlprefix  [0]{URL }%
\providecommand \Eprint [0]{\href }%
\providecommand \doibase [0]{http://dx.doi.org/}%
\providecommand \selectlanguage [0]{\@gobble}%
\providecommand \bibinfo  [0]{\@secondoftwo}%
\providecommand \bibfield  [0]{\@secondoftwo}%
\providecommand \translation [1]{[#1]}%
\providecommand \BibitemOpen [0]{}%
\providecommand \bibitemStop [0]{}%
\providecommand \bibitemNoStop [0]{.\EOS\space}%
\providecommand \EOS [0]{\spacefactor3000\relax}%
\providecommand \BibitemShut  [1]{\csname bibitem#1\endcsname}%
\let\auto@bib@innerbib\@empty
\bibitem [{\citenamefont {Hofstadter}(1976)}]{Hofstadter:prb76}%
  \BibitemOpen
  \bibfield  {author} {\bibinfo {author} {\bibfnamefont {D.~R.}\ \bibnamefont
  {Hofstadter}},\ }\href {\doibase 10.1103/PhysRevB.14.2239} {\bibfield
  {journal} {\bibinfo  {journal} {Phys. Rev. B}\ }\textbf {\bibinfo {volume}
  {14}},\ \bibinfo {pages} {2239} (\bibinfo {year} {1976})}\BibitemShut
  {NoStop}%
\bibitem [{\citenamefont {Li}\ \emph {et~al.}(2011)\citenamefont {Li},
  \citenamefont {Wang},\ and\ \citenamefont {Gong}}]{LiJ:jpcm11}%
  \BibitemOpen
  \bibfield  {author} {\bibinfo {author} {\bibfnamefont {J.}~\bibnamefont
  {Li}}, \bibinfo {author} {\bibfnamefont {Y.-F.}\ \bibnamefont {Wang}}, \ and\
  \bibinfo {author} {\bibfnamefont {C.-D.}\ \bibnamefont {Gong}},\ }\href
  {\doibase 10.1088/0953-8984/23/15/156002} {\bibfield  {journal} {\bibinfo
  {journal} {J. Phys. Condens. Matter}\ }\textbf {\bibinfo {volume} {23}},\
  \bibinfo {pages} {156002} (\bibinfo {year} {2011})}\BibitemShut {NoStop}%
\bibitem [{\citenamefont {Kuhl}\ and\ \citenamefont
  {St{\"{o}}ckmann}(1998)}]{Kuhl:prl98}%
  \BibitemOpen
  \bibfield  {author} {\bibinfo {author} {\bibfnamefont {U.}~\bibnamefont
  {Kuhl}}\ and\ \bibinfo {author} {\bibfnamefont {H.~J.}\ \bibnamefont
  {St{\"{o}}ckmann}},\ }\href {\doibase 10.1103/PhysRevLett.80.3232} {\bibfield
   {journal} {\bibinfo  {journal} {Phys. Rev. Lett.}\ }\textbf {\bibinfo
  {volume} {80}},\ \bibinfo {pages} {3232} (\bibinfo {year}
  {1998})}\BibitemShut {NoStop}%
\bibitem [{\citenamefont {Dean}\ \emph {et~al.}(2013)\citenamefont {Dean},
  \citenamefont {Wang}, \citenamefont {Maher}, \citenamefont {Forsythe},
  \citenamefont {Ghahari}, \citenamefont {Gao}, \citenamefont {Katoch},
  \citenamefont {Ishigami}, \citenamefont {Moon}, \citenamefont {Koshino},
  \citenamefont {Taniguchi}, \citenamefont {Watanabe}, \citenamefont {Shepard},
  \citenamefont {Hone},\ and\ \citenamefont {Kim}}]{Dean:nat13}%
  \BibitemOpen
  \bibfield  {author} {\bibinfo {author} {\bibfnamefont {C.~R.}\ \bibnamefont
  {Dean}}, \bibinfo {author} {\bibfnamefont {L.}~\bibnamefont {Wang}}, \bibinfo
  {author} {\bibfnamefont {P.}~\bibnamefont {Maher}}, \bibinfo {author}
  {\bibfnamefont {C.}~\bibnamefont {Forsythe}}, \bibinfo {author}
  {\bibfnamefont {F.}~\bibnamefont {Ghahari}}, \bibinfo {author} {\bibfnamefont
  {Y.}~\bibnamefont {Gao}}, \bibinfo {author} {\bibfnamefont {J.}~\bibnamefont
  {Katoch}}, \bibinfo {author} {\bibfnamefont {M.}~\bibnamefont {Ishigami}},
  \bibinfo {author} {\bibfnamefont {P.}~\bibnamefont {Moon}}, \bibinfo {author}
  {\bibfnamefont {M.}~\bibnamefont {Koshino}}, \bibinfo {author} {\bibfnamefont
  {T.}~\bibnamefont {Taniguchi}}, \bibinfo {author} {\bibfnamefont
  {K.}~\bibnamefont {Watanabe}}, \bibinfo {author} {\bibfnamefont {K.~L.}\
  \bibnamefont {Shepard}}, \bibinfo {author} {\bibfnamefont {J.}~\bibnamefont
  {Hone}}, \ and\ \bibinfo {author} {\bibfnamefont {P.}~\bibnamefont {Kim}},\
  }\href {\doibase 10.1038/nature12186} {\bibfield  {journal} {\bibinfo
  {journal} {Nature}\ }\textbf {\bibinfo {volume} {497}},\ \bibinfo {pages}
  {598} (\bibinfo {year} {2013})}\BibitemShut {NoStop}%
\bibitem [{\citenamefont {Ponomarenko}\ \emph {et~al.}(2013)\citenamefont
  {Ponomarenko}, \citenamefont {Gorbachev}, \citenamefont {Yu}, \citenamefont
  {Elias}, \citenamefont {Jalil}, \citenamefont {Patel}, \citenamefont
  {Mishchenko}, \citenamefont {Mayorov}, \citenamefont {Woods}, \citenamefont
  {Wallbank}, \citenamefont {Mucha-Kruczynski}, \citenamefont {Piot},
  \citenamefont {Potemski}, \citenamefont {Grigorieva}, \citenamefont
  {Novoselov}, \citenamefont {Guinea}, \citenamefont {Fal'ko},\ and\
  \citenamefont {Geim}}]{Ponomarenko:nat13}%
  \BibitemOpen
  \bibfield  {author} {\bibinfo {author} {\bibfnamefont {L.~A.}\ \bibnamefont
  {Ponomarenko}}, \bibinfo {author} {\bibfnamefont {R.~V.}\ \bibnamefont
  {Gorbachev}}, \bibinfo {author} {\bibfnamefont {G.~L.}\ \bibnamefont {Yu}},
  \bibinfo {author} {\bibfnamefont {D.~C.}\ \bibnamefont {Elias}}, \bibinfo
  {author} {\bibfnamefont {R.}~\bibnamefont {Jalil}}, \bibinfo {author}
  {\bibfnamefont {A.~A.}\ \bibnamefont {Patel}}, \bibinfo {author}
  {\bibfnamefont {A.}~\bibnamefont {Mishchenko}}, \bibinfo {author}
  {\bibfnamefont {A.~S.}\ \bibnamefont {Mayorov}}, \bibinfo {author}
  {\bibfnamefont {C.~R.}\ \bibnamefont {Woods}}, \bibinfo {author}
  {\bibfnamefont {J.~R.}\ \bibnamefont {Wallbank}}, \bibinfo {author}
  {\bibfnamefont {M.}~\bibnamefont {Mucha-Kruczynski}}, \bibinfo {author}
  {\bibfnamefont {B.~A.}\ \bibnamefont {Piot}}, \bibinfo {author}
  {\bibfnamefont {M.}~\bibnamefont {Potemski}}, \bibinfo {author}
  {\bibfnamefont {I.~V.}\ \bibnamefont {Grigorieva}}, \bibinfo {author}
  {\bibfnamefont {K.~S.}\ \bibnamefont {Novoselov}}, \bibinfo {author}
  {\bibfnamefont {F.}~\bibnamefont {Guinea}}, \bibinfo {author} {\bibfnamefont
  {V.~I.}\ \bibnamefont {Fal'ko}}, \ and\ \bibinfo {author} {\bibfnamefont
  {A.~K.}\ \bibnamefont {Geim}},\ }\href {\doibase 10.1038/nature12187}
  {\bibfield  {journal} {\bibinfo  {journal} {Nature}\ }\textbf {\bibinfo
  {volume} {497}},\ \bibinfo {pages} {594} (\bibinfo {year}
  {2013})}\BibitemShut {NoStop}%
\bibitem [{\citenamefont {Hunt}\ \emph {et~al.}(2013)\citenamefont {Hunt},
  \citenamefont {Sanchez-Yamagishi}, \citenamefont {Young}, \citenamefont
  {Yankowitz}, \citenamefont {LeRoy}, \citenamefont {Watanabe}, \citenamefont
  {Taniguchi}, \citenamefont {Moon}, \citenamefont {Koshino}, \citenamefont
  {Jarillo-Herrero},\ and\ \citenamefont {Ashoori}}]{Hunt:sci13}%
  \BibitemOpen
  \bibfield  {author} {\bibinfo {author} {\bibfnamefont {B.}~\bibnamefont
  {Hunt}}, \bibinfo {author} {\bibfnamefont {J.~D.}\ \bibnamefont
  {Sanchez-Yamagishi}}, \bibinfo {author} {\bibfnamefont {A.~F.}\ \bibnamefont
  {Young}}, \bibinfo {author} {\bibfnamefont {M.}~\bibnamefont {Yankowitz}},
  \bibinfo {author} {\bibfnamefont {B.~J.}\ \bibnamefont {LeRoy}}, \bibinfo
  {author} {\bibfnamefont {K.}~\bibnamefont {Watanabe}}, \bibinfo {author}
  {\bibfnamefont {T.}~\bibnamefont {Taniguchi}}, \bibinfo {author}
  {\bibfnamefont {P.}~\bibnamefont {Moon}}, \bibinfo {author} {\bibfnamefont
  {M.}~\bibnamefont {Koshino}}, \bibinfo {author} {\bibfnamefont
  {P.}~\bibnamefont {Jarillo-Herrero}}, \ and\ \bibinfo {author} {\bibfnamefont
  {R.~C.}\ \bibnamefont {Ashoori}},\ }\href {\doibase 10.1126/science.1237240}
  {\bibfield  {journal} {\bibinfo  {journal} {Science}\ }\textbf {\bibinfo
  {volume} {340}},\ \bibinfo {pages} {1427} (\bibinfo {year}
  {2013})}\BibitemShut {NoStop}%
\bibitem [{\citenamefont {Roushan}\ \emph {et~al.}(2017)\citenamefont
  {Roushan}, \citenamefont {Neill}, \citenamefont {Tangpanitanon},
  \citenamefont {Bastidas}, \citenamefont {Megrant}, \citenamefont {Barends},
  \citenamefont {Chen}, \citenamefont {Chen}, \citenamefont {Chiaro},
  \citenamefont {Dunsworth}, \citenamefont {Fowler}, \citenamefont {Foxen},
  \citenamefont {Giustina}, \citenamefont {Jeffrey}, \citenamefont {Kelly},
  \citenamefont {Lucero}, \citenamefont {Mutus}, \citenamefont {Neeley},
  \citenamefont {Quintana}, \citenamefont {Sank}, \citenamefont {Vainsencher},
  \citenamefont {Wenner}, \citenamefont {White}, \citenamefont {Neven},
  \citenamefont {Angelakis},\ and\ \citenamefont {Martinis}}]{Roushan:sci17}%
  \BibitemOpen
  \bibfield  {author} {\bibinfo {author} {\bibfnamefont {P.}~\bibnamefont
  {Roushan}}, \bibinfo {author} {\bibfnamefont {C.}~\bibnamefont {Neill}},
  \bibinfo {author} {\bibfnamefont {J.}~\bibnamefont {Tangpanitanon}}, \bibinfo
  {author} {\bibfnamefont {V.~M.}\ \bibnamefont {Bastidas}}, \bibinfo {author}
  {\bibfnamefont {A.}~\bibnamefont {Megrant}}, \bibinfo {author} {\bibfnamefont
  {R.}~\bibnamefont {Barends}}, \bibinfo {author} {\bibfnamefont
  {Y.}~\bibnamefont {Chen}}, \bibinfo {author} {\bibfnamefont {Z.}~\bibnamefont
  {Chen}}, \bibinfo {author} {\bibfnamefont {B.}~\bibnamefont {Chiaro}},
  \bibinfo {author} {\bibfnamefont {A.}~\bibnamefont {Dunsworth}}, \bibinfo
  {author} {\bibfnamefont {A.}~\bibnamefont {Fowler}}, \bibinfo {author}
  {\bibfnamefont {B.}~\bibnamefont {Foxen}}, \bibinfo {author} {\bibfnamefont
  {M.}~\bibnamefont {Giustina}}, \bibinfo {author} {\bibfnamefont
  {E.}~\bibnamefont {Jeffrey}}, \bibinfo {author} {\bibfnamefont
  {J.}~\bibnamefont {Kelly}}, \bibinfo {author} {\bibfnamefont
  {E.}~\bibnamefont {Lucero}}, \bibinfo {author} {\bibfnamefont
  {J.}~\bibnamefont {Mutus}}, \bibinfo {author} {\bibfnamefont
  {M.}~\bibnamefont {Neeley}}, \bibinfo {author} {\bibfnamefont
  {C.}~\bibnamefont {Quintana}}, \bibinfo {author} {\bibfnamefont
  {D.}~\bibnamefont {Sank}}, \bibinfo {author} {\bibfnamefont {A.}~\bibnamefont
  {Vainsencher}}, \bibinfo {author} {\bibfnamefont {J.}~\bibnamefont {Wenner}},
  \bibinfo {author} {\bibfnamefont {T.}~\bibnamefont {White}}, \bibinfo
  {author} {\bibfnamefont {H.}~\bibnamefont {Neven}}, \bibinfo {author}
  {\bibfnamefont {D.~G.}\ \bibnamefont {Angelakis}}, \ and\ \bibinfo {author}
  {\bibfnamefont {J.}~\bibnamefont {Martinis}},\ }\href {\doibase
  10.1126/science.aao1401} {\bibfield  {journal} {\bibinfo  {journal}
  {Science}\ }\textbf {\bibinfo {volume} {358}},\ \bibinfo {pages} {1175}
  (\bibinfo {year} {2017})}\BibitemShut {NoStop}%
\bibitem [{\citenamefont {Kitagawa}\ \emph {et~al.}(2010)\citenamefont
  {Kitagawa}, \citenamefont {Berg}, \citenamefont {Rudner},\ and\ \citenamefont
  {Demler}}]{Kitagawa:prb10}%
  \BibitemOpen
  \bibfield  {author} {\bibinfo {author} {\bibfnamefont {T.}~\bibnamefont
  {Kitagawa}}, \bibinfo {author} {\bibfnamefont {E.}~\bibnamefont {Berg}},
  \bibinfo {author} {\bibfnamefont {M.}~\bibnamefont {Rudner}}, \ and\ \bibinfo
  {author} {\bibfnamefont {E.}~\bibnamefont {Demler}},\ }\href {\doibase
  10.1103/PhysRevB.82.235114} {\bibfield  {journal} {\bibinfo  {journal} {Phys.
  Rev. B}\ }\textbf {\bibinfo {volume} {82}},\ \bibinfo {pages} {235114}
  (\bibinfo {year} {2010})}\BibitemShut {NoStop}%
\bibitem [{\citenamefont {Rudner}\ \emph {et~al.}(2013)\citenamefont {Rudner},
  \citenamefont {Lindner}, \citenamefont {Berg},\ and\ \citenamefont
  {Levin}}]{Rudner:prx13}%
  \BibitemOpen
  \bibfield  {author} {\bibinfo {author} {\bibfnamefont {M.~S.}\ \bibnamefont
  {Rudner}}, \bibinfo {author} {\bibfnamefont {N.~H.}\ \bibnamefont {Lindner}},
  \bibinfo {author} {\bibfnamefont {E.}~\bibnamefont {Berg}}, \ and\ \bibinfo
  {author} {\bibfnamefont {M.}~\bibnamefont {Levin}},\ }\href {\doibase
  10.1103/PhysRevX.3.031005} {\bibfield  {journal} {\bibinfo  {journal} {Phys.
  Rev. X}\ }\textbf {\bibinfo {volume} {3}},\ \bibinfo {pages} {031005}
  (\bibinfo {year} {2013})}\BibitemShut {NoStop}%
\bibitem [{\citenamefont {Katan}\ and\ \citenamefont
  {Podolsky}(2013)}]{Katan:prl13}%
  \BibitemOpen
  \bibfield  {author} {\bibinfo {author} {\bibfnamefont {Y.~T.}\ \bibnamefont
  {Katan}}\ and\ \bibinfo {author} {\bibfnamefont {D.}~\bibnamefont
  {Podolsky}},\ }\href {\doibase 10.1103/PhysRevLett.110.016802} {\bibfield
  {journal} {\bibinfo  {journal} {Phys. Rev. Lett.}\ }\textbf {\bibinfo
  {volume} {110}},\ \bibinfo {pages} {016802} (\bibinfo {year}
  {2013})}\BibitemShut {NoStop}%
\bibitem [{\citenamefont {Lindner}\ \emph {et~al.}(2013)\citenamefont
  {Lindner}, \citenamefont {Bergman}, \citenamefont {Refael},\ and\
  \citenamefont {Galitski}}]{Lindner:prb13}%
  \BibitemOpen
  \bibfield  {author} {\bibinfo {author} {\bibfnamefont {N.~H.}\ \bibnamefont
  {Lindner}}, \bibinfo {author} {\bibfnamefont {D.~L.}\ \bibnamefont
  {Bergman}}, \bibinfo {author} {\bibfnamefont {G.}~\bibnamefont {Refael}}, \
  and\ \bibinfo {author} {\bibfnamefont {V.}~\bibnamefont {Galitski}},\ }\href
  {\doibase 10.1103/PhysRevB.87.235131} {\bibfield  {journal} {\bibinfo
  {journal} {Phys. Rev. B}\ }\textbf {\bibinfo {volume} {87}},\ \bibinfo
  {pages} {235131} (\bibinfo {year} {2013})},\ \Eprint
  {http://arxiv.org/abs/1111.4518} {arXiv:1111.4518} \BibitemShut {NoStop}%
\bibitem [{\citenamefont {D{\'{o}}ra}\ \emph {et~al.}(2012)\citenamefont
  {D{\'{o}}ra}, \citenamefont {Cayssol}, \citenamefont {Simon},\ and\
  \citenamefont {Moessner}}]{Dora:prl12}%
  \BibitemOpen
  \bibfield  {author} {\bibinfo {author} {\bibfnamefont {B.}~\bibnamefont
  {D{\'{o}}ra}}, \bibinfo {author} {\bibfnamefont {J.}~\bibnamefont {Cayssol}},
  \bibinfo {author} {\bibfnamefont {F.}~\bibnamefont {Simon}}, \ and\ \bibinfo
  {author} {\bibfnamefont {R.}~\bibnamefont {Moessner}},\ }\href {\doibase
  10.1103/PhysRevLett.108.056602} {\bibfield  {journal} {\bibinfo  {journal}
  {Phys. Rev. Lett.}\ }\textbf {\bibinfo {volume} {108}},\ \bibinfo {pages}
  {056602} (\bibinfo {year} {2012})}\BibitemShut {NoStop}%
\bibitem [{\citenamefont {Inoue}\ and\ \citenamefont
  {Tanaka}(2010)}]{Inoue:prl10}%
  \BibitemOpen
  \bibfield  {author} {\bibinfo {author} {\bibfnamefont {J.-i.}\ \bibnamefont
  {Inoue}}\ and\ \bibinfo {author} {\bibfnamefont {A.}~\bibnamefont {Tanaka}},\
  }\href {\doibase 10.1103/PhysRevLett.105.017401} {\bibfield  {journal}
  {\bibinfo  {journal} {Phys. Rev. Lett.}\ }\textbf {\bibinfo {volume} {105}},\
  \bibinfo {pages} {017401} (\bibinfo {year} {2010})}\BibitemShut {NoStop}%
\bibitem [{\citenamefont {Cayssol}\ \emph {et~al.}(2013)\citenamefont
  {Cayssol}, \citenamefont {D{\'{o}}ra}, \citenamefont {Simon},\ and\
  \citenamefont {Moessner}}]{Cayssol:pssr13}%
  \BibitemOpen
  \bibfield  {author} {\bibinfo {author} {\bibfnamefont {J.}~\bibnamefont
  {Cayssol}}, \bibinfo {author} {\bibfnamefont {B.}~\bibnamefont {D{\'{o}}ra}},
  \bibinfo {author} {\bibfnamefont {F.}~\bibnamefont {Simon}}, \ and\ \bibinfo
  {author} {\bibfnamefont {R.}~\bibnamefont {Moessner}},\ }\href {\doibase
  10.1002/pssr.201206451} {\bibfield  {journal} {\bibinfo  {journal} {Phys.
  Status Solidi - Rapid Res. Lett.}\ }\textbf {\bibinfo {volume} {7}},\
  \bibinfo {pages} {101} (\bibinfo {year} {2013})},\ \Eprint
  {http://arxiv.org/abs/1211.5623} {arXiv:1211.5623} \BibitemShut {NoStop}%
\bibitem [{\citenamefont {Kitagawa}\ \emph {et~al.}(2011)\citenamefont
  {Kitagawa}, \citenamefont {Oka}, \citenamefont {Brataas}, \citenamefont
  {Fu},\ and\ \citenamefont {Demler}}]{Kitagawa:prb11}%
  \BibitemOpen
  \bibfield  {author} {\bibinfo {author} {\bibfnamefont {T.}~\bibnamefont
  {Kitagawa}}, \bibinfo {author} {\bibfnamefont {T.}~\bibnamefont {Oka}},
  \bibinfo {author} {\bibfnamefont {A.}~\bibnamefont {Brataas}}, \bibinfo
  {author} {\bibfnamefont {L.}~\bibnamefont {Fu}}, \ and\ \bibinfo {author}
  {\bibfnamefont {E.}~\bibnamefont {Demler}},\ }\href {\doibase
  10.1103/PhysRevB.84.235108} {\bibfield  {journal} {\bibinfo  {journal} {Phys.
  Rev. B}\ }\textbf {\bibinfo {volume} {84}} (\bibinfo {year} {2011}),\
  10.1103/PhysRevB.84.235108}\BibitemShut {NoStop}%
\bibitem [{\citenamefont {Iadecola}\ \emph {et~al.}(2013)\citenamefont
  {Iadecola}, \citenamefont {Campbell}, \citenamefont {Chamon}, \citenamefont
  {Hou}, \citenamefont {Jackiw}, \citenamefont {Pi},\ and\ \citenamefont
  {Kusminskiy}}]{Iadecola:prl13}%
  \BibitemOpen
  \bibfield  {author} {\bibinfo {author} {\bibfnamefont {T.}~\bibnamefont
  {Iadecola}}, \bibinfo {author} {\bibfnamefont {D.}~\bibnamefont {Campbell}},
  \bibinfo {author} {\bibfnamefont {C.}~\bibnamefont {Chamon}}, \bibinfo
  {author} {\bibfnamefont {C.-Y.}\ \bibnamefont {Hou}}, \bibinfo {author}
  {\bibfnamefont {R.}~\bibnamefont {Jackiw}}, \bibinfo {author} {\bibfnamefont
  {S.-Y.}\ \bibnamefont {Pi}}, \ and\ \bibinfo {author} {\bibfnamefont {S.~V.}\
  \bibnamefont {Kusminskiy}},\ }\href {\doibase 10.1103/PhysRevLett.110.176603}
  {\bibfield  {journal} {\bibinfo  {journal} {Phys. Rev. Lett.}\ }\textbf
  {\bibinfo {volume} {110}},\ \bibinfo {pages} {176603} (\bibinfo {year}
  {2013})}\BibitemShut {NoStop}%
\bibitem [{\citenamefont {Ezawa}(2013)}]{Ezawa:prl13}%
  \BibitemOpen
  \bibfield  {author} {\bibinfo {author} {\bibfnamefont {M.}~\bibnamefont
  {Ezawa}},\ }\href {\doibase 10.1103/PhysRevLett.110.026603} {\bibfield
  {journal} {\bibinfo  {journal} {Phys. Rev. Lett.}\ }\textbf {\bibinfo
  {volume} {110}},\ \bibinfo {pages} {026603} (\bibinfo {year}
  {2013})}\BibitemShut {NoStop}%
\bibitem [{\citenamefont {Kemper}\ \emph {et~al.}(2013)\citenamefont {Kemper},
  \citenamefont {Sentef}, \citenamefont {Moritz}, \citenamefont {Kao},
  \citenamefont {Shen}, \citenamefont {Freericks},\ and\ \citenamefont
  {Devereaux}}]{Kemper:prb13}%
  \BibitemOpen
  \bibfield  {author} {\bibinfo {author} {\bibfnamefont {A.~F.}\ \bibnamefont
  {Kemper}}, \bibinfo {author} {\bibfnamefont {M.}~\bibnamefont {Sentef}},
  \bibinfo {author} {\bibfnamefont {B.}~\bibnamefont {Moritz}}, \bibinfo
  {author} {\bibfnamefont {C.~C.}\ \bibnamefont {Kao}}, \bibinfo {author}
  {\bibfnamefont {Z.~X.}\ \bibnamefont {Shen}}, \bibinfo {author}
  {\bibfnamefont {J.~K.}\ \bibnamefont {Freericks}}, \ and\ \bibinfo {author}
  {\bibfnamefont {T.~P.}\ \bibnamefont {Devereaux}},\ }\href {\doibase
  10.1103/PhysRevB.87.235139} {\bibfield  {journal} {\bibinfo  {journal} {Phys.
  Rev. B}\ }\textbf {\bibinfo {volume} {87}},\ \bibinfo {pages} {235139}
  (\bibinfo {year} {2013})}\BibitemShut {NoStop}%
\bibitem [{\citenamefont {Rechtsman}\ \emph {et~al.}(2013)\citenamefont
  {Rechtsman}, \citenamefont {Zeuner}, \citenamefont {Plotnik}, \citenamefont
  {Lumer}, \citenamefont {Podolsky}, \citenamefont {Dreisow}, \citenamefont
  {Nolte}, \citenamefont {Segev},\ and\ \citenamefont
  {Szameit}}]{Rechtsman:nat13}%
  \BibitemOpen
  \bibfield  {author} {\bibinfo {author} {\bibfnamefont {M.~C.}\ \bibnamefont
  {Rechtsman}}, \bibinfo {author} {\bibfnamefont {J.~M.}\ \bibnamefont
  {Zeuner}}, \bibinfo {author} {\bibfnamefont {Y.}~\bibnamefont {Plotnik}},
  \bibinfo {author} {\bibfnamefont {Y.}~\bibnamefont {Lumer}}, \bibinfo
  {author} {\bibfnamefont {D.}~\bibnamefont {Podolsky}}, \bibinfo {author}
  {\bibfnamefont {F.}~\bibnamefont {Dreisow}}, \bibinfo {author} {\bibfnamefont
  {S.}~\bibnamefont {Nolte}}, \bibinfo {author} {\bibfnamefont
  {M.}~\bibnamefont {Segev}}, \ and\ \bibinfo {author} {\bibfnamefont
  {A.}~\bibnamefont {Szameit}},\ }\href {\doibase 10.1038/nature12066}
  {\bibfield  {journal} {\bibinfo  {journal} {Nature}\ }\textbf {\bibinfo
  {volume} {496}},\ \bibinfo {pages} {196} (\bibinfo {year}
  {2013})}\BibitemShut {NoStop}%
\bibitem [{\citenamefont {Du}\ \emph {et~al.}(2017)\citenamefont {Du},
  \citenamefont {Zhou},\ and\ \citenamefont {Fiete}}]{DuL:prb17a}%
  \BibitemOpen
  \bibfield  {author} {\bibinfo {author} {\bibfnamefont {L.}~\bibnamefont
  {Du}}, \bibinfo {author} {\bibfnamefont {X.}~\bibnamefont {Zhou}}, \ and\
  \bibinfo {author} {\bibfnamefont {G.~A.}\ \bibnamefont {Fiete}},\ }\href
  {\doibase 10.1103/PhysRevB.95.035136} {\bibfield  {journal} {\bibinfo
  {journal} {Phys. Rev. B}\ }\textbf {\bibinfo {volume} {95}},\ \bibinfo
  {pages} {035136} (\bibinfo {year} {2017})}\BibitemShut {NoStop}%
\bibitem [{\citenamefont {Chen}\ \emph {et~al.}(2018)\citenamefont {Chen},
  \citenamefont {Du},\ and\ \citenamefont {Fiete}}]{ChenQ:prb18}%
  \BibitemOpen
  \bibfield  {author} {\bibinfo {author} {\bibfnamefont {Q.}~\bibnamefont
  {Chen}}, \bibinfo {author} {\bibfnamefont {L.}~\bibnamefont {Du}}, \ and\
  \bibinfo {author} {\bibfnamefont {G.~A.}\ \bibnamefont {Fiete}},\ }\href
  {\doibase 10.1103/PhysRevB.97.035422} {\bibfield  {journal} {\bibinfo
  {journal} {Phys. Rev. B}\ }\textbf {\bibinfo {volume} {97}},\ \bibinfo
  {pages} {035422} (\bibinfo {year} {2018})}\BibitemShut {NoStop}%
\bibitem [{\citenamefont {Jotzu}\ \emph {et~al.}(2014)\citenamefont {Jotzu},
  \citenamefont {Messer}, \citenamefont {Desbuquois}, \citenamefont {Lebrat},
  \citenamefont {Uehlinger}, \citenamefont {Greif},\ and\ \citenamefont
  {Esslinger}}]{Jotzu:nat14}%
  \BibitemOpen
  \bibfield  {author} {\bibinfo {author} {\bibfnamefont {G.}~\bibnamefont
  {Jotzu}}, \bibinfo {author} {\bibfnamefont {M.}~\bibnamefont {Messer}},
  \bibinfo {author} {\bibfnamefont {R.}~\bibnamefont {Desbuquois}}, \bibinfo
  {author} {\bibfnamefont {M.}~\bibnamefont {Lebrat}}, \bibinfo {author}
  {\bibfnamefont {T.}~\bibnamefont {Uehlinger}}, \bibinfo {author}
  {\bibfnamefont {D.}~\bibnamefont {Greif}}, \ and\ \bibinfo {author}
  {\bibfnamefont {T.}~\bibnamefont {Esslinger}},\ }\href {\doibase
  10.1038/nature13915} {\bibfield  {journal} {\bibinfo  {journal} {Nature}\
  }\textbf {\bibinfo {volume} {515}},\ \bibinfo {pages} {237} (\bibinfo {year}
  {2014})}\BibitemShut {NoStop}%
\bibitem [{\citenamefont {Bilitewski}\ and\ \citenamefont
  {Cooper}(2015)}]{Bilitewski:pra15}%
  \BibitemOpen
  \bibfield  {author} {\bibinfo {author} {\bibfnamefont {T.}~\bibnamefont
  {Bilitewski}}\ and\ \bibinfo {author} {\bibfnamefont {N.~R.}\ \bibnamefont
  {Cooper}},\ }\href {\doibase 10.1103/PhysRevA.91.063611} {\bibfield
  {journal} {\bibinfo  {journal} {Phys. Rev. A}\ }\textbf {\bibinfo {volume}
  {91}},\ \bibinfo {pages} {063611} (\bibinfo {year} {2015})}\BibitemShut
  {NoStop}%
\bibitem [{\citenamefont {Fregoso}\ \emph {et~al.}(2013)\citenamefont
  {Fregoso}, \citenamefont {Wang}, \citenamefont {Gedik},\ and\ \citenamefont
  {Galitski}}]{Fregoso:prb13}%
  \BibitemOpen
  \bibfield  {author} {\bibinfo {author} {\bibfnamefont {B.~M.}\ \bibnamefont
  {Fregoso}}, \bibinfo {author} {\bibfnamefont {Y.~H.}\ \bibnamefont {Wang}},
  \bibinfo {author} {\bibfnamefont {N.}~\bibnamefont {Gedik}}, \ and\ \bibinfo
  {author} {\bibfnamefont {V.}~\bibnamefont {Galitski}},\ }\href {\doibase
  10.1103/PhysRevB.88.155129} {\bibfield  {journal} {\bibinfo  {journal} {Phys.
  Rev. B}\ }\textbf {\bibinfo {volume} {88}},\ \bibinfo {pages} {155129}
  (\bibinfo {year} {2013})}\BibitemShut {NoStop}%
\bibitem [{\citenamefont {Sentef}\ \emph {et~al.}(2015)\citenamefont {Sentef},
  \citenamefont {Claassen}, \citenamefont {Kemper}, \citenamefont {Moritz},
  \citenamefont {Oka}, \citenamefont {Freericks},\ and\ \citenamefont
  {Devereaux}}]{Sentef:nc15}%
  \BibitemOpen
  \bibfield  {author} {\bibinfo {author} {\bibfnamefont {M.}~\bibnamefont
  {Sentef}}, \bibinfo {author} {\bibfnamefont {M.}~\bibnamefont {Claassen}},
  \bibinfo {author} {\bibfnamefont {A.}~\bibnamefont {Kemper}}, \bibinfo
  {author} {\bibfnamefont {B.}~\bibnamefont {Moritz}}, \bibinfo {author}
  {\bibfnamefont {T.}~\bibnamefont {Oka}}, \bibinfo {author} {\bibfnamefont
  {J.}~\bibnamefont {Freericks}}, \ and\ \bibinfo {author} {\bibfnamefont
  {T.}~\bibnamefont {Devereaux}},\ }\href {\doibase 10.1038/ncomms8047}
  {\bibfield  {journal} {\bibinfo  {journal} {Nat. Commun.}\ }\textbf {\bibinfo
  {volume} {6}},\ \bibinfo {pages} {7047} (\bibinfo {year} {2015})}\BibitemShut
  {NoStop}%
\bibitem [{\citenamefont {Wang}\ \emph {et~al.}(2013)\citenamefont {Wang},
  \citenamefont {Steinberg}, \citenamefont {Jarillo-Herrero},\ and\
  \citenamefont {Gedik}}]{WangY:sci13}%
  \BibitemOpen
  \bibfield  {author} {\bibinfo {author} {\bibfnamefont {Y.~H.}\ \bibnamefont
  {Wang}}, \bibinfo {author} {\bibfnamefont {H.}~\bibnamefont {Steinberg}},
  \bibinfo {author} {\bibfnamefont {P.}~\bibnamefont {Jarillo-Herrero}}, \ and\
  \bibinfo {author} {\bibfnamefont {N.}~\bibnamefont {Gedik}},\ }\href
  {\doibase 10.1126/science.1239834} {\bibfield  {journal} {\bibinfo  {journal}
  {Science}\ }\textbf {\bibinfo {volume} {342}},\ \bibinfo {pages} {453}
  (\bibinfo {year} {2013})}\BibitemShut {NoStop}%
\bibitem [{\citenamefont {Mahmood}\ \emph {et~al.}(2016)\citenamefont
  {Mahmood}, \citenamefont {Chan}, \citenamefont {Alpichshev}, \citenamefont
  {Gardner}, \citenamefont {Lee}, \citenamefont {Lee},\ and\ \citenamefont
  {Gedik}}]{Mahmood:np16}%
  \BibitemOpen
  \bibfield  {author} {\bibinfo {author} {\bibfnamefont {F.}~\bibnamefont
  {Mahmood}}, \bibinfo {author} {\bibfnamefont {C.-K.}\ \bibnamefont {Chan}},
  \bibinfo {author} {\bibfnamefont {Z.}~\bibnamefont {Alpichshev}}, \bibinfo
  {author} {\bibfnamefont {D.}~\bibnamefont {Gardner}}, \bibinfo {author}
  {\bibfnamefont {Y.}~\bibnamefont {Lee}}, \bibinfo {author} {\bibfnamefont
  {P.~A.}\ \bibnamefont {Lee}}, \ and\ \bibinfo {author} {\bibfnamefont
  {N.}~\bibnamefont {Gedik}},\ }\href {\doibase 10.1038/nphys3609} {\bibfield
  {journal} {\bibinfo  {journal} {Nat. Phys.}\ }\textbf {\bibinfo {volume}
  {12}},\ \bibinfo {pages} {306} (\bibinfo {year} {2016})}\BibitemShut
  {NoStop}%
\bibitem [{\citenamefont {Calvo}\ \emph {et~al.}(2015)\citenamefont {Calvo},
  \citenamefont {{Foa Torres}}, \citenamefont {Perez-Piskunow}, \citenamefont
  {Balseiro},\ and\ \citenamefont {Usaj}}]{Calvo:prb15}%
  \BibitemOpen
  \bibfield  {author} {\bibinfo {author} {\bibfnamefont {H.~L.}\ \bibnamefont
  {Calvo}}, \bibinfo {author} {\bibfnamefont {L.~E.~F.}\ \bibnamefont {{Foa
  Torres}}}, \bibinfo {author} {\bibfnamefont {P.~M.}\ \bibnamefont
  {Perez-Piskunow}}, \bibinfo {author} {\bibfnamefont {C.~A.}\ \bibnamefont
  {Balseiro}}, \ and\ \bibinfo {author} {\bibfnamefont {G.}~\bibnamefont
  {Usaj}},\ }\href {\doibase 10.1103/PhysRevB.91.241404} {\bibfield  {journal}
  {\bibinfo  {journal} {Phys. Rev. B}\ }\textbf {\bibinfo {volume} {91}},\
  \bibinfo {pages} {241404} (\bibinfo {year} {2015})}\BibitemShut {NoStop}%
\bibitem [{\citenamefont {{Dal Lago}}\ \emph {et~al.}(2015)\citenamefont {{Dal
  Lago}}, \citenamefont {Atala},\ and\ \citenamefont {{Foa
  Torres}}}]{Lago:pra15}%
  \BibitemOpen
  \bibfield  {author} {\bibinfo {author} {\bibfnamefont {V.}~\bibnamefont {{Dal
  Lago}}}, \bibinfo {author} {\bibfnamefont {M.}~\bibnamefont {Atala}}, \ and\
  \bibinfo {author} {\bibfnamefont {L.~E.~F.}\ \bibnamefont {{Foa Torres}}},\
  }\href {\doibase 10.1103/PhysRevA.92.023624} {\bibfield  {journal} {\bibinfo
  {journal} {Phys. Rev. A}\ }\textbf {\bibinfo {volume} {92}},\ \bibinfo
  {pages} {023624} (\bibinfo {year} {2015})}\BibitemShut {NoStop}%
\bibitem [{\citenamefont {Perez-Piskunow}\ \emph {et~al.}(2015)\citenamefont
  {Perez-Piskunow}, \citenamefont {{Foa Torres}},\ and\ \citenamefont
  {Usaj}}]{Perez-Piskunow:pra15}%
  \BibitemOpen
  \bibfield  {author} {\bibinfo {author} {\bibfnamefont {P.~M.}\ \bibnamefont
  {Perez-Piskunow}}, \bibinfo {author} {\bibfnamefont {L.~E.~F.}\ \bibnamefont
  {{Foa Torres}}}, \ and\ \bibinfo {author} {\bibfnamefont {G.}~\bibnamefont
  {Usaj}},\ }\href {\doibase 10.1103/PhysRevA.91.043625} {\bibfield  {journal}
  {\bibinfo  {journal} {Phys. Rev. A}\ }\textbf {\bibinfo {volume} {91}},\
  \bibinfo {pages} {043625} (\bibinfo {year} {2015})}\BibitemShut {NoStop}%
\bibitem [{\citenamefont {Perez-Piskunow}\ \emph {et~al.}(2014)\citenamefont
  {Perez-Piskunow}, \citenamefont {Usaj}, \citenamefont {Balseiro},\ and\
  \citenamefont {Torres}}]{Perez-Piskunow:prb14}%
  \BibitemOpen
  \bibfield  {author} {\bibinfo {author} {\bibfnamefont {P.~M.}\ \bibnamefont
  {Perez-Piskunow}}, \bibinfo {author} {\bibfnamefont {G.}~\bibnamefont
  {Usaj}}, \bibinfo {author} {\bibfnamefont {C.~A.}\ \bibnamefont {Balseiro}},
  \ and\ \bibinfo {author} {\bibfnamefont {L.~E. F.~F.}\ \bibnamefont
  {Torres}},\ }\href {\doibase 10.1103/PhysRevB.89.121401} {\bibfield
  {journal} {\bibinfo  {journal} {Phys. Rev. B}\ }\textbf {\bibinfo {volume}
  {89}},\ \bibinfo {pages} {121401} (\bibinfo {year} {2014})}\BibitemShut
  {NoStop}%
\bibitem [{\citenamefont {Lababidi}\ \emph {et~al.}(2014)\citenamefont
  {Lababidi}, \citenamefont {Satija},\ and\ \citenamefont
  {Zhao}}]{Lababidi:prl14}%
  \BibitemOpen
  \bibfield  {author} {\bibinfo {author} {\bibfnamefont {M.}~\bibnamefont
  {Lababidi}}, \bibinfo {author} {\bibfnamefont {I.~I.}\ \bibnamefont
  {Satija}}, \ and\ \bibinfo {author} {\bibfnamefont {E.}~\bibnamefont
  {Zhao}},\ }\href {\doibase 10.1103/PhysRevLett.112.026805} {\bibfield
  {journal} {\bibinfo  {journal} {Phys. Rev. Lett.}\ }\textbf {\bibinfo
  {volume} {112}},\ \bibinfo {pages} {026805} (\bibinfo {year}
  {2014})}\BibitemShut {NoStop}%
\bibitem [{\citenamefont {Zhou}\ \emph {et~al.}(2014)\citenamefont {Zhou},
  \citenamefont {Satija},\ and\ \citenamefont {Zhao}}]{ZhouZ:prb14}%
  \BibitemOpen
  \bibfield  {author} {\bibinfo {author} {\bibfnamefont {Z.}~\bibnamefont
  {Zhou}}, \bibinfo {author} {\bibfnamefont {I.~I.}\ \bibnamefont {Satija}}, \
  and\ \bibinfo {author} {\bibfnamefont {E.}~\bibnamefont {Zhao}},\ }\href
  {\doibase 10.1103/PhysRevB.90.205108} {\bibfield  {journal} {\bibinfo
  {journal} {Phys. Rev. B}\ }\textbf {\bibinfo {volume} {90}},\ \bibinfo
  {pages} {205108} (\bibinfo {year} {2014})}\BibitemShut {NoStop}%
\bibitem [{\citenamefont {Wackerl}\ and\ \citenamefont
  {Schliemann}(2018)}]{Wackerl:arXiv18}%
  \BibitemOpen
  \bibfield  {author} {\bibinfo {author} {\bibfnamefont {M.}~\bibnamefont
  {Wackerl}}\ and\ \bibinfo {author} {\bibfnamefont {J.}~\bibnamefont
  {Schliemann}},\ }\href {http://arxiv.org/abs/1802.01369} {\bibfield
  {journal} {\bibinfo  {journal} {arXiv}\ } (\bibinfo {year} {2018})},\ \Eprint
  {http://arxiv.org/abs/1802.01369} {arXiv:1802.01369} \BibitemShut {NoStop}%
\bibitem [{\citenamefont {Kooi}\ \emph {et~al.}(2018)\citenamefont {Kooi},
  \citenamefont {Quelle}, \citenamefont {Beugeling},\ and\ \citenamefont
  {Smith}}]{Kooi:arXiv18}%
  \BibitemOpen
  \bibfield  {author} {\bibinfo {author} {\bibfnamefont {S.~H.}\ \bibnamefont
  {Kooi}}, \bibinfo {author} {\bibfnamefont {A.}~\bibnamefont {Quelle}},
  \bibinfo {author} {\bibfnamefont {W.}~\bibnamefont {Beugeling}}, \ and\
  \bibinfo {author} {\bibfnamefont {C.~M.}\ \bibnamefont {Smith}},\ }\href
  {http://arxiv.org/abs/1803.04791} {\bibfield  {journal} {\bibinfo  {journal}
  {arXiv}\ } (\bibinfo {year} {2018})},\ \Eprint
  {http://arxiv.org/abs/1803.04791} {arXiv:1803.04791} \BibitemShut {NoStop}%
\bibitem [{\citenamefont {Osadchy}\ and\ \citenamefont
  {Avron}(2001)}]{Osadchy:jmp01}%
  \BibitemOpen
  \bibfield  {author} {\bibinfo {author} {\bibfnamefont {D.}~\bibnamefont
  {Osadchy}}\ and\ \bibinfo {author} {\bibfnamefont {J.~E.}\ \bibnamefont
  {Avron}},\ }\href {\doibase 10.1063/1.1412464} {\bibfield  {journal}
  {\bibinfo  {journal} {J. Math. Phys.}\ }\textbf {\bibinfo {volume} {42}},\
  \bibinfo {pages} {5665} (\bibinfo {year} {2001})}\BibitemShut {NoStop}%
\bibitem [{\citenamefont {Goldman}(2009)}]{Goldman:jpb09}%
  \BibitemOpen
  \bibfield  {author} {\bibinfo {author} {\bibfnamefont {N.}~\bibnamefont
  {Goldman}},\ }\href {\doibase 10.1088/0953-4075/42/5/055302} {\bibfield
  {journal} {\bibinfo  {journal} {J. Phys. B At. Mol. Opt. Phys.}\ }\textbf
  {\bibinfo {volume} {42}},\ \bibinfo {pages} {055302} (\bibinfo {year}
  {2009})}\BibitemShut {NoStop}%
\bibitem [{\citenamefont {Fukui}\ \emph {et~al.}(2005)\citenamefont {Fukui},
  \citenamefont {Hatsugai},\ and\ \citenamefont {Suzuki}}]{Fukui:jpsj05}%
  \BibitemOpen
  \bibfield  {author} {\bibinfo {author} {\bibfnamefont {T.}~\bibnamefont
  {Fukui}}, \bibinfo {author} {\bibfnamefont {Y.}~\bibnamefont {Hatsugai}}, \
  and\ \bibinfo {author} {\bibfnamefont {H.}~\bibnamefont {Suzuki}},\ }\href
  {\doibase 10.1143/JPSJ.74.1674} {\bibfield  {journal} {\bibinfo  {journal}
  {J. Phys. Soc. Japan}\ }\textbf {\bibinfo {volume} {74}},\ \bibinfo {pages}
  {1674} (\bibinfo {year} {2005})}\BibitemShut {NoStop}%
\bibitem [{\citenamefont {Dana}\ \emph {et~al.}(1985)\citenamefont {Dana},
  \citenamefont {Avron},\ and\ \citenamefont {Zak}}]{Dana:jpc85}%
  \BibitemOpen
  \bibfield  {author} {\bibinfo {author} {\bibfnamefont {I.}~\bibnamefont
  {Dana}}, \bibinfo {author} {\bibfnamefont {Y.}~\bibnamefont {Avron}}, \ and\
  \bibinfo {author} {\bibfnamefont {J.}~\bibnamefont {Zak}},\ }\href
  {http://stacks.iop.org/0022-3719/18/i=22/a=004} {\bibfield  {journal}
  {\bibinfo  {journal} {Journal of Physics C: Solid State Physics}\ }\textbf
  {\bibinfo {volume} {18}},\ \bibinfo {pages} {L679} (\bibinfo {year}
  {1985})}\BibitemShut {NoStop}%
\end{thebibliography}%
\end{document}